\documentclass[fleqn,usenatbib]{mnras}

\usepackage{newtxtext,newtxmath}

\usepackage[T1]{fontenc}
\usepackage{ae,aecompl}

\usepackage{graphicx}	
\usepackage{amsmath}	
\usepackage{multicol} 
\usepackage{multirow}
\usepackage{comment}
\usepackage{array}
\usepackage{soul}
\hypersetup{colorlinks=true,citecolor=blue,linkcolor=purple,filecolor=black,runcolor=black,breaklinks=true}
\usepackage{etoolbox}
\usepackage{ae,aecompl}
\usepackage[usenames]{color}
\usepackage{hyperref}
\usepackage{threeparttable}
\usepackage{mathrsfs}
\usepackage{scalerel}
\usepackage{longtable,booktabs,threeparttablex}
\usepackage{float}

\definecolor{purple}{RGB}{160,32,240}

\definecolor{red}{RGB}{225,50,50}

\definecolor{addchange}{RGB}{215,25,25}

\definecolor{removechange}{RGB}{25,25,215}

\newcommand{\HST}{\emph{HST}}
\newcommand{\JWST}{\emph{JWST}}
\newcommand{\Spitzer}{\emph{Spitzer}}
\newcommand{\Muv}{\ensuremath{\mathrm{M}_{\mathrm{UV}}^{ }}}

\newcommand{\Lya}{\ensuremath{\mathrm{Ly}\alpha}}

\newcommand{\xiion}{\ensuremath{\xi_{\mathrm{ion}}^{\ast}}}

\newcommand{\logxiion}{\ensuremath{\log[\xi_{\mathrm{ion}}^{\ast} / (\mathrm{erg}^{-1}\ \mathrm{Hz})]}}

\newcommand{\fesc}{\ensuremath{f_{\mathrm{esc}}}}

\newcommand{\OIIIHb}{[OIII]+H\ensuremath{\beta}}

\newcommand{\Msol}{\ensuremath{M_{\odot}}}
\newcommand{\Mstar}{\ensuremath{M_{\ast}}}
\newcommand{\logMstar}{\ensuremath{\log\left(M_{\ast}/M_{\odot}\right)}}

\newcommand{\MstarCSFH}{\ensuremath{M_{\ast\mathrm{,CSFH}}}}
\newcommand{\ageCSFH}{\ensuremath{\mathrm{age_{CSFH}}}}
\newcommand{\sSFRCSFH}{\ensuremath{\mathrm{sSFR_{CSFH}}}}

\newcolumntype{P}[1]{>{\centering\arraybackslash}p{#1}}
\newcommand\Tstrut{\rule{0pt}{2.6ex}}         
\newcommand\Bstrut{\rule[-1.2ex]{0pt}{0pt}}   

\title[The Nature of UV-faint $z\sim7-8$ Galaxies]{A JWST/NIRCam Study of Key Contributors to Reionization: The Star-forming and Ionizing Properties of UV-faint $z\sim7-8$ Galaxies}

\author[R. Endsley et al.]{Ryan Endsley$^{1,2}$\thanks{E-mail: ryan.endsley@austin.utexas.edu},
Daniel P. Stark$^{1}$, 
Lily Whitler$^{1}$,
Michael W. Topping$^{1}$,
Zuyi Chen$^{1}$,
Adele Plat$^{1}$,
\newauthor
John Chisholm$^{2}$,
St\'ephane Charlot$^{3}$
\\
$^{1}$Steward Observatory, University of Arizona, 933 N Cherry Ave, Tucson, AZ 85721 USA\\
$^{2}$Department of Astronomy, University of Texas, Austin, TX 78712, USA\\
$^{3}$Sorbonne Universit\'e, UPMC-CNRS, UMR7095, Institut d'Astrophysique de Paris, F-75014, Paris, France
}

\date{Accepted XXX. Received YYY; in original form ZZZ}

\pubyear{2023}

\begin{document}
\label{firstpage}
\pagerange{\pageref{firstpage}--\pageref{lastpage}}
\maketitle


\begin{abstract}

\noindent
\textit{Spitzer}/IRAC imaging has revealed that the brightest $z\sim7-8$ galaxies often exhibit young ages and strong nebular line emission, hinting at high ionizing efficiency among early galaxies. However, IRAC's limited sensitivity has long hindered efforts to study the fainter, more numerous population often thought largely responsible for reionization. Here we use CEERS \textit{JWST}/NIRCam data to characterize 116 UV-faint (median M$_\mathrm{UV}=-19.5$) $z\sim6.5-8$ galaxies. The SEDs are typically dominated by young ($\sim$10--50 Myr), low-mass ($M_\ast\sim10^8\ M_\odot$) stellar populations, and we find no need for extremely high stellar masses ($\sim10^{11}\ M_\odot$). Considering previous studies of UV-bright (M$_\mathrm{UV}\sim-22$) $z\sim7-8$ galaxies, we find evidence for a strong (5--10$\times$) increase in specific star formation rate toward lower luminosities (median sSFR=103 Gyr$^{-1}$ in CEERS). The larger sSFRs imply a more dominant contribution from OB stars in the relatively numerous UV-faint population, perhaps suggesting that these galaxies are very efficient ionizing agents (median $\xi_\mathrm{ion}=10^{25.7}$ erg$^{-1}$ Hz). In spite of the much larger sSFRs, we find little increase in [OIII]$+$H$\beta$ EWs towards fainter M$_\mathrm{UV}$ (median $\approx$780 $\mathrm{\mathring{A}}$). If confirmed, this may indicate that a substantial fraction of our CEERS galaxies possess extremely low metallicities ($\lesssim$3\% $Z_\odot$) where [OIII] emission is suppressed. Alternatively, high ionizing photon escape fractions or bursty star formation histories can also weaken the nebular lines in a subset of our CEERS galaxies. While the majority of our objects are very blue (median $\beta=-2.0$), we identify a significant tail of very dusty galaxies ($\beta\sim-1$) at $\approx$0.5 $L_\mathrm{UV}^\ast$ which may contribute significantly to the $z\sim7-8$ star formation rate density.
\end{abstract}




\section{Introduction} \label{sec:intro}

The epoch of hydrogen reionization reflects when galaxy formation began impacting the large-scale ionization state of the Universe \citep{Dayal2018,Robertson2022}.
Current observational constraints imply that nearly all hydrogen atoms in the intergalactic medium were reionized by $z\simeq5.3$ \citep{Bosman2022}, with this process approximately halfway complete by $z\sim7-8$ \citep[e.g.][]{Davies2018,Mason2018_IGMneutralFrac,Jung2020,Planck2020,Wang2020,Yang2020_Poniuaena,Goto2021}.
Given the rapidly declining quasar luminosity function at $z>3$, it has long been thought that star-forming galaxies are likely the dominant contributors to reionization \citep[e.g.][]{Madau1999,Ciardi2000}, motivating much effort to characterize these systems in detail.

Thanks to a large collection of dedicated surveys, our initial understanding of galaxies present during the reionization era has emerged over the past two decades.
In the mid-to-late 2000s, the first small samples of Lyman-break galaxies at $z\gtrsim7$ were identified in deep near-infrared images taken with ground-based facilities as well as the NICMOS camera on the \textit{Hubble Space Telescope} (\HST{}; \citealt{Bouwens2004,Bouwens2008,Kneib2004,Yan2004,Mannucci2007,Stanway2008}).
Our census of these early galaxies advanced tremendously in the 2010s following the installation of the Wide Field Camera 3 (WFC3) on \HST{}, affording an $\approx$40$\times$ increased near-IR survey efficiency relative to NICMOS.
To date, several deep \HST{} imaging surveys have been conducted in both blind and gravitationally lensed fields \citep{Bouwens2011_HUDF,Grogin2011,Koekemoer2011,Trenti2011,Windhorst2011,Illingworth2013,Lotz2017,Coe2019} resulting in the discovery of approximately 1000 Lyman-break galaxies at $z\sim7-10$ with intrinsic UV continuum luminosities spanning three orders of magnitude ($-22 \lesssim \Muv{} \lesssim -14$; e.g. \citealt{McLure2013,Atek2015a,Bouwens2015_LF,Bouwens2021_LF,Finkelstein2015_LF,Finkelstein2022,McLeod2016,Livermore2017,Ishigaki2018,Oesch2018_z10LF,Salmon2020}).
The derived UV continuum luminosity functions at $z\sim7-10$ have revealed a characteristic UV luminosity ($L_\mathrm{UV}^\ast$) equivalent to $\mathrm{M}_\mathrm{UV}^\ast \sim -21$ with a steep faint-end slope ($\alpha \sim -2$) that steadily steepens with increasing redshift \citep[e.g.][]{Finkelstein2015_LF,Ishigaki2018,Bouwens2021_LF}, indicating a highly dominant population of UV-faint ($\Muv{} \gtrsim -20$) galaxies in the early Universe.
From the WFC3 data, the UV continuum slopes of $z\sim7-9$ galaxies were typically found to be very blue with $-2.5 \lesssim \beta \lesssim -2$ (where $f_\lambda \propto \lambda^\beta$; e.g. \citealt{McLure2011,Dunlop2012,Finkelstein2012,Rogers2013,Bouwens2014_beta,Bhatawdekar2021}), suggesting little-to-no dust attenuation in most systems.
However, the rest-UV regime probed by WFC3 lacked the information necessary to characterize the stellar populations dominating the observed spectral energy distributions (SEDs).

Prior to the launch of the \textit{James Webb Space Telescope} (\JWST{}), our primary insights into the ages and stellar masses of $z\gtrsim7$ galaxies have come from the Infrared Array Camera (IRAC) on board the \textit{Spitzer Space Telescope} which could probe the rest-optical emission of reionization-era systems.
Early studies frequently found strong photometric excesses between the near-infrared and IRAC 3--5$\mu$m data of faint $z\sim5-7$ galaxies, which was at the time interpreted as extremely strong Balmer breaks indicative of very old ($\sim$300 Myr) and massive ($\Mstar{} \sim 10^{10-11} \Msol{}$) stellar populations \citep[e.g.][]{Egami2005,Eyles2007,Stark2009,Gonzalez2010}.
Not only did this imply extremely rapid stellar mass growth at $z>8$, but also little evolution in galaxy specific star formation rates (sSFRs) towards earlier epochs in contrast to expectations \citep[e.g.][]{Weinmann2011,Dave2011,Dave2012,Dayal2012}. 

By the early 2010s it was realized that high equivalent width (EW) rest-optical nebular emission lines (e.g. H$\alpha$, H$\beta$, and [OIII]$\lambda$4959,5007) can significantly boost the IRAC broadband photometry from high-redshift galaxies \citep[e.g.][]{Schaerer2009,Schaerer2010,Ono2010,Labbe2013,Stark2013_NebEmission,deBarros2014_NebEmission,Gonzalez2014,Smit2014}.
This meant that the IRAC photometric excesses seen in many faint $z\gtrsim5$ galaxies could potentially be explained by either a very strong Balmer break, or very high EW lines powered by a much younger ($\lesssim$50 Myr), lower-mass ($\Mstar{} \sim 10^{7-8} \Msol{}$) stellar population.
Because higher EW nebular lines are typically associated with more efficient production of hydrogen ionizing photons \citep[e.g.][]{Chevallard2018_z0,Tang2019}, these alternative solutions could have substantial implications for our understanding of how galaxies contributed to reionization.
Unfortunately, breaking this degeneracy proved to be very challenging given that IRAC data suffered from inadequate sensitivity ($m\sim26$ at 5$\sigma$ in the deepest fields), strong source confusion due to its broad point spread function (FWHM$\approx$2\arcsec{}), and limited sampling of the rest-optical SED with only two filters at 3--5$\mu$m.
The sensitivity limitations of IRAC alone frequently forced studies of faint ($< L_\mathrm{UV}^\ast$) reionization-era galaxy SEDs to resort to stacking, leaving much uncertainty on the demographics of ages, stellar masses, sSFRs, and nebular line strengths among this population.
While deep IRAC data in lensing fields enabled studies to overcome some of these issues \citep[e.g.][]{Huang2016_SURFSUP,Strait2020}, the numbers of UV-faint $z\gtrsim7$ galaxies detected at high IRAC S/N remained small, necessitating further work to characterize this population.

The successful development, launch, and commissioning of \JWST{} now provides the opportunity to completely transform our understanding of typical ($<\!L_\mathrm{UV}^\ast$; $m\gtrsim26$) reionization-era galaxies.
In this work, we specifically focus on utilizing the greatly improved 1--5$\mu$m imaging capabilities of \JWST{}/NIRCam. 
By reaching 5$\sigma$ depths of $m\sim29$ with just 1 hour integrations, NIRCam is $\sim$30,000$\times$ more sensitive than IRAC at 3--5$\mu$m ($m\approx26$ in 120 hours; e.g. \citealt{Labbe2013,Oesch2013_z9LF}).
Such a dramatic gain in sensitivity would alone revolutionize our ability to characterize the rest-optical SEDs of individual faint reionization-era galaxies.
Thankfully, \JWST{}/NIRCam also provides two other key improvements over \Spitzer{}/IRAC enabling us to push our understanding of $z\gtrsim7$ systems even further.
The large mirror size of \JWST{} results in a $\approx$15-fold improvement in angular resolution over \Spitzer{} at 3--5$\mu$m, greatly mitigating past source confusion issues that would cause large numbers of viable $z\gtrsim7$ candidates to be discarded from rest-optical analyses.
Finally, NIRCam provides a large suite of medium- and broad-band filters at $\approx$3--5$\mu$m which not only improves photometric redshift precision (since the strongest nebular lines impact different bands at different redshifts), but also greatly alleviates degeneracies in interpreting photometric excesses as arising from strong Balmer breaks versus very high-EW nebular lines.
Consequently, NIRCam data is able to substantially clarify not only the star-forming properties of UV-faint reionization-era galaxies, but their ionizing properties as well.

Our primary goal in this work is to begin developing a much clearer picture of the physical properties (e.g. ages, stellar masses, sSFRs, nebular line EWs, and dust optical depths) of typical (sub-$L_\mathrm{UV}^\ast$) reionization-era galaxies by characterizing their rest-UV$+$optical SEDs in detail with \JWST{}/NIRCam imaging data.
To do so, we utilize data from the Cosmic Evolution Early Release Science (CEERS\footnote{\url{https://ceers.github.io/}}; \citealt{Finkelstein2017_CEERS,Finkelstein2023_CEERS}) survey which provides deep NIRCam imaging in seven bands from 1--5$\mu$m, including four bands at $\approx$3--5$\mu$m.
We focus our attention on UV-selected Lyman-break galaxies at $z\sim6.5-8$ where the 3--5$\mu$m filter set of CEERS/NIRCam affords rich insight into both the strength of nebular emission lines as well as the underlying rest-optical continuum.

The structure of this paper is as follows.
We begin by describing the imaging data, source extraction, photometric calculations, sample selection, and photoionization modelling in \S\ref{sec:sec2}.
We then present the variety of NIRCam SEDs measured among our sample of 116 Lyman-break $z\sim6.5-8$ galaxies, discussing implications for demographics in ages, stellar masses, and sSFRs of typical reionization-era systems (\S\ref{sec:SEDs}).
Next we discuss two unexpected yet seemingly important classes of sub-$L_\mathrm{UV}^\ast$ $z\sim6.5-8$ galaxies that are newly revealed with the NIRCam data (\S\ref{sec:weakOIII}--\ref{sec:dusty}).
We then combine the results from our faint $z\sim6.5-8$ galaxy sample with previous wide-area results on the UV-bright $z\sim7-8$ population (where IRAC data was sufficiently sensitive) to begin investigating any UV luminosity dependence on sSFR or nebular line EWs (\S\ref{sec:sSFRandEW}).
Finally, we briefly discuss whether we find evidence of extremely massive ($M_\ast \sim 10^{11}\ M_\odot$) galaxies in our sample, comparing with recent \JWST{}/NIRCam studies (\S\ref{sec:stellarMasses}).

Throughout this paper, we quote all magnitudes in the AB system, assume a \citet{Chabrier2003} initial mass function (IMF) with limits of 0.1--300 \Msol{}, and adopt a flat $\Lambda$CDM cosmology with parameters $h=0.7$, $\Omega_\mathrm{M}=0.3$, and $\Omega_\mathrm{\Lambda}=0.7$. We provide a catalog listing coordinates, magnitudes, and inferred physical properties among our sample.

\section{Imaging Data and Sample Selection} \label{sec:sec2}

We begin this section by describing the CEERS \JWST{}/NIRCam imaging data over the Extended Groth Strip (EGS) field, as well as the overlapping optical \HST{}/ACS imaging utilized for our dropout selection (\S\ref{sec:imaging}).
Our source extraction and photometric calculations are then detailed in \S\ref{sec:photometry}, wherein we also describe and demonstrate a custom algorithm that removes contaminating flux from nearby objects to help produce reliable photometry.
The Lyman-break criteria used to select our final $z\sim6.5-8$ galaxy sample are described in \S\ref{sec:sample} where we also provide a brief description of the photometric properties of the sample.
Finally, we describe the star-forming photoionization model SED fitting procedure used to infer the physical properties for each of our galaxies in \S\ref{sec:beagle}.

\subsection{\JWST{} and \HST{} Imaging in EGS} \label{sec:imaging}

The CEERS \JWST{}/NIRCam imaging utilized in this work were taken in June 2022 and cover $\approx$40 arcmin$^2$ over the EGS field in six broadband filters (F115W, F150W, F200W, F277W, F356W, and F444W) as well as one medium-band filter (F410M).
In the redshift window we aim to select galaxies ($z\sim6.5-8$), the rest-optical SED is probed by at least three of these NIRCam filters (F356W, F410M, and F444W) while F277W additionally covers part of the rest-optical regime at $z\lesssim7.3$.
The availability of F410M photometry in particular delivers key information on whether any photometric excesses in these bands are likely driven by a strong Balmer break or contamination from high-EW nebular line emission.

Before detailing our NIRCam image reduction and photometric calculations, we first briefly describe the optical \HST{} imaging that we employ to identify Lyman-break $z\sim6.5-8$ galaxies across the CEERS NIRCam footprint. 
The EGS field possesses deep imaging in three optical filters (F435W, F606W, and F814W) taken with the Advanced Camera for Surveys (ACS) as part of the following \HST{} surveys: the All-Wavelength Extended Groth Strip International Survey (AEGIS; \citealt{davis2007}), the Cosmic Assembly Near-infrared Deep Extragalactic Legacy Survey (CANDELS; \citealt{Grogin2011,Koekemoer2011}), and the Ultraviolet Imaging of the Cosmic Assembly Near-infrared Deep Extragalactic Legacy Survey Fields (UVCANDELS\footnote{\url{https://archive.stsci.edu/hlsp/uvcandels}}; PI: Teplitz).
All ACS data from these surveys were processed into co-added calibrated mosaics using the \textsc{grizli} software \citep{grizli2022} as part of the Complete Hubble Archive for Galaxy Evolution (CHArGE) project (Kokorev in prep.).
The CHArGE data products also include mosaics in various \HST{}/WFC3 bands compiled from archival data. 
All mosaics from CHArGE are aligned to the \textit{Gaia} astrometric frame with the ACS (WFC3) mosaics set to a pixel scale of 40 mas/pixel (80 mas/pixel).

Co-added mosaics for each NIRCam band are produced by first downloading the uncalibrated ({\tt *\_rate.fits}) NIRCam exposures from the MAST Portal.\footnote{\url{https://mast.stsci.edu/portal/Mashup/Clients/Mast/Portal.html}}
These data are processed through the \JWST{} Science Calibration Pipeline\footnote{\url{https://jwst-pipeline.readthedocs.io/en/latest/index.html}} where we group exposures by NIRCam filter, pointing, and module combination to achieve precise astrometric alignment (see \citealt{Chen2023}).
We utilize the most recent NIRCam photometric calibration reference files, (first released as {\tt jwst\_0942.pmap} in early October 2022), and have verified that the total F150W and F200W magnitudes of bright isolated objects in our final mosaics are on average consistent with the WFC3/F160W and WIRCam/$K_s$ magnitudes, respectively, reported in previous EGS catalogs \citep{Skelton2014,Stefanon2017_EGS}.
After individual co-added images ({\tt *\_i2d.fits}) are output from the stage 3 step of the pipeline, we align each (one per filter, pointing, and module combination) to the \textit{Gaia} frame.
Because the surface density of \textit{Gaia} stars is very low across the CEERS footprint ($\lesssim$3 per module area), direct alignment is not feasible.
Instead we utilize \textsc{tweakreg}\footnote{\url{https://drizzlepac.readthedocs.io/en/latest/tweakreg.html}} to align each co-added image to the CHArGE WFC3/F160W mosaic (which is pre-registered to \textit{Gaia}), yielding precise relative alignment (RMS$\approx$6--15 mas).
With consistent astrometry across the individual co-adds, we then resample all images for each filter onto a fixed grid with pixel scale of 30 mas/pixel.
Background subtraction is then performed on each mosaic using the \textsc{sep} package \citep{Barbary2016_sep}.

As a last step to producing our final mosaics, we mitigate $1/f$ noise.
To do so, we first run \textsc{Source Extractor} \citep{Bertin1996} on the F200W mosaic.
The output source catalog is used to mask objects in the uncalibrated NIRCam exposures before performing a row-by-row background subtraction on these images, largely following the algorithm\footnote{\url{https://www.stsci.edu/jwst/science-planning/proposal-planning-toolbox/simulated-data}} produced by the CEERS team to minimize $1/f$ noise \citep{Bagley2023}.
The background-subtracted uncalibrated files are then re-processed as described in the above paragraph to produce our final mosaics.

When performing aperture photometry on the \JWST{} and \HST{} data, we must account for the fact that the width of the point spread function (PSF) varies by a factor of $\approx$4 across the different mosaics (FWHM$\approx$45--180 mas).
One option is to convolve all mosaics to the lowest-resolution PSF (WFC3/160W) prior to calculating photometry.
However, doing so would significantly weaken the sensitivity in the shortest-wavelength bands which probe the Lyman-alpha break at our redshifts of interest ($z\sim6.5-8$).
To mitigate this effect, we choose to convolve all long-wavelength
(LW) NIRCam mosaics (F277W, F356W, F410M, F444W) to the PSF of WFC3/F160W while convolving all ACS and short-wavelength (SW) NIRCam mosaics (F115W, F150W, F200W) to the PSF of ACS/F814W.
As described below, we introduce a scaling factor to the aperture size used for the LW vs. ACS+SW photometry to account for the differing PSFs.
For each mosaic, we generate an empirical PSF by stacking the normalized surface brightness profiles of at least five stars within the mosaic footprint that were visually inspected to be unsaturated and isolated.
Convolution kernels are then created using the \textsc{photutils} package \citep{Bradley2022_photutils}, where we verify that the encircled energy distributions after convolution agree with that of the target PSF (WFC3/F160W for LW and ACS/F814W for ACS+SW) to within $\lesssim$2\% accuracy at radii $>$0.05\arcsec{}.

\begin{figure*}
\includegraphics{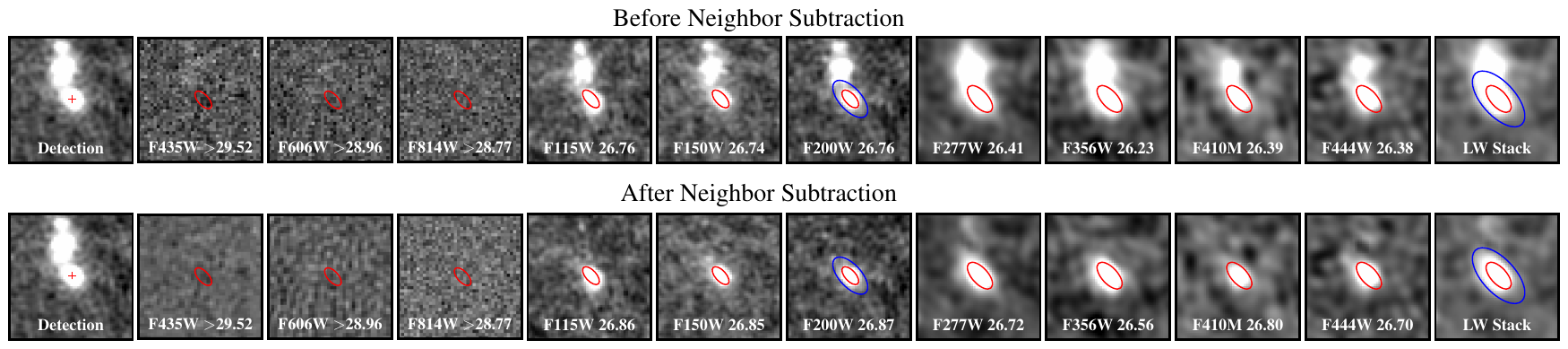}
\caption{An illustration of how our neighbor subtraction algorithm improves  photometric measurements for objects with close neighbors. The $z\sim6.5-8$ galaxy of interest is marked with a red cross in the leftmost panels which show the F150W$+$F200W \textsc{Source Extractor} detection image. The $k=1.2$ apertures are shown in red for each band, and we also show the $k=2.5$ apertures in blue for the F200W and inverse variance-weighted LW stack image. After subtracting the surface brightness profiles of neighboring sources contaminating these apertures, we find that the ACS$+$SW fluxes decrease by $\approx$0.1 mag while the LW fluxes decrease by $\approx$0.3--0.4 mag given the broader PSF and larger aperture size. Each postage stamp is 1.6$\times$1.6 arcsec$^2$ and we show the measured AB magnitude for each band at the bottom, where we quote 2$\sigma$ limits for the ACS bands.}
\label{fig:neighborSubtraction}
\end{figure*}

\subsection{Source Extraction and Photometry} \label{sec:photometry}

Objects within the CEERS footprint are identified by running \textsc{Source Extractor} on an inverse-variance weighted stack of the PSF-convolved F150W and F200W mosaics, yielding a total of $\approx$51,000 sources. 
We stack the F150W and F200W mosaics for the detection image as we aim to perform a rest-UV selection of $z\sim6.5-8$ in this work.
The resulting catalog was also utilized in \citet{Whitler2023_z10} to select galaxies at $z\sim8.5-11$ which appear as F115W dropouts.

The photometry of each object identified by \textsc{source extractor} is calculated following procedures adopted in past \HST{}-based analyses of $z\gtrsim6$ galaxies \citep[e.g.][]{Oesch2013_z9LF,Bouwens2015_LF,Bouwens2021_LF,Finkelstein2015_LF,Finkelstein2022,Endsley2021_OIII}.
We begin by measuring the flux contained in elliptical apertures with a \citet{Kron1980} factor of $k=1.2$ which is expected to result in approximately maximum signal to noise \citep{Finkelstein2022}.
The associated photometric error in each band is calculated as the standard deviation of flux contained in equally-sized Kron apertures randomly distributed in nearby empty regions of the sky as determined from a \textsc{Source Extractor} segmentation map specific to each band.
Because we have homogenized the ACS$+$SW and LW mosaics to different PSFs (ACS/F814W and WFC3/F160W, respectively; \S\ref{sec:imaging}), we multiply the aperture size used for the LW bands by a factor of 1.5 which reflects the typical F444W vs. F200W size ratio of twelve UV-bright $z\sim6-8$ galaxy candidates measured from the PSF-homogenized mosaics \citep{Chen2023}.

The $k=1.2$ photometry (and their associated errors) are corrected to total values by first multiplying by the ratio of flux measured in $k=2.5$ vs. $k=1.2$ apertures.
This correction factor is computed from the PSF-homogenized F200W mosaic for the ACS$+$SW photometry given that our $z\sim6.5-8$ selection requires a $>$5$\sigma$ detection in F200W (\S\ref{sec:sample}).
Since we do not explicitly require a detection in any individual LW band, the correction factor for the photometry in these four bands are computed from an inverse variance-weighted stack of their mosaics to maximize S/N.
Finally we correct for flux outside the $k=2.5$ apertures using our constructed ACS/F814W and WFC3/F160W mosaics as well as their reported encircled energy distributions at large radii ($>$1\arcsec{}).

While our PSF-homogenized mosaics have very high angular resolution (FWHM$<$0.2\arcsec{}), it is still possible for our photometric apertures to be contaminated by considerable flux from neighboring objects.
This can introduce significant offsets to not only the total magnitudes of our sources, but also the colors between the SW and LW NIRCam photometry given the different adopted PSF homogenization.
To address this concern, we have developed a neighbor subtraction algorithm that is largely based off techniques previously developed for IRAC deconfusion \citep[e.g.][]{Wuyts2007,Labbe2010,McLeod2016,Song2016,Endsley2021_OIII}.
In short, this algorithm fits a S{\'e}rsic profile (convolved with the appropriate PSF) to each neighboring object identified in the \textsc{Source Extractor} catalog within a 4$\times$4 arcsec$^2$ region.
The fit is performed on the non PSF-homogenized image of each band independently, adopting the source parameters output by \textsc{Source Extractor} as starting values and iterating to find the best-fitting solution via a least-squares optimization.
Once the best-fitting surface brightness parameters are determined for each of the non-homogenized images, we then subtract the flux of neighboring objects from the corresponding PSF-homogenized images by convolving the inferred S{\'e}rsic profiles with the F814W or F160W PSF for the ACS$+$SW and LW data, respectively.

We demonstrate the performance and benefit of our neighbor subtraction algorithm in Fig. \ref{fig:neighborSubtraction}.
In this example, significant flux from a bright neighboring low-redshift object is contaminating the Kron apertures for one of our $z\sim6.5-8$ galaxies, thereby artificially boosting the measured fluxes.
Because the LW photometry is performed with a larger aperture size and a broader homogenized PSF, the impact of the contaminating flux is systematically stronger for the LW bands than ACS$+$SW.
For this particular example, subtracting the surface brightness profiles of neighboring objects decreases the measured SW fluxes by $\approx$0.1 mag while the LW fluxes decrease by $\approx$0.3--0.4 mags.
Not correcting for this contaminating flux would therefore result in seemingly stronger SW-to-LW colors, potentially giving the false impression of a stronger Balmer break and hence higher stellar mass and older age.
Given these potential effects, we produce two photometric catalogs -- one that implements the neighbor subtraction algorithm and one without neighbor subtraction -- and perform our $z\sim6.5-8$ galaxy selection on each.
This ensures that we identify the maximum number of viable reionization-era galaxies, as some may have biased colors or ACS detections in the original catalog due to contaminating flux from neighbors.

\subsection{Selection of Lyman-break \texorpdfstring{$\mathbf{z\sim6.5-8}$}{z ~ 6.5 - 8} Galaxies} \label{sec:sample}

We perform a Lyman-break selection to identify galaxies at $z\sim6.5-8$ across the June 2022 CEERS footprint. 
At $z>6.5$, the intergalactic medium is highly opaque to both Lyman-series ($\lambda_\mathrm{rest} = 912-1216$ \AA{}) and Lyman-continuum (LyC; $\lambda_\mathrm{rest} < 912$ \AA{}) emission \citep[e.g.][]{Inoue2014}. 
Accordingly, we expect galaxies at $z>6.5$ to show negligible emission in bands that lie blueward of $\approx$9100 \AA{}, resulting in a strong dropout in the \HST{}/ACS photometry relative to that of NIRCam F115W.
With this in mind, we develop the follow initial cuts to identify Lyman-break $z\sim6.5-8$ galaxies within CEERS:
\begin{enumerate}
    \item S/N$<$2 in F435W, F606W, and F814W
    \item F606W$-$F115W$>$1.7 and F814W$-$F115W$>$1.7
    \item F115W$-$F200W$<$1.0
    \item F814W$-$F115W$>$(F115W$-$F200W)$+$1.5
\end{enumerate}
Here, the flux densities in the dropout bands (F606W and F814W) are set to their 1$\sigma$ upper limit in cases of non-detections (S/N$<$1), consistent with the approach of previous works \citep[e.g.][]{Bouwens2015_LF,Endsley2021_OIII}.
The first two selection criteria above ensure that the photometry are consistent with expectations of a Lyman-alpha break at $z\gtrsim6.5$. 
The third and fourth criteria limit our sample to objects that show a much redder F814W$-$F115W color relative to F115W$-$F200W, thereby requiring the presence of a sharp break while still allowing $z\sim6.5-8$ objects with very red rest-UV slopes ($-1.5 \lesssim \beta \lesssim -0.3$) to enter our sample (see \S\ref{sec:dusty}).
The overlapping CEERS+ACS area over which we are able to apply our selection is $\approx$31 arcmin$^2$.

In addition to the above cuts, we add the following selection criteria to help remove spurious detections and low-redshift interlopers.
First, we require that each selected source be detected at S/N$>$5 in F200W, S/N$>$3 in F150W, as well as S/N$>$3 in at least two LW NIRCam bands.
Second, we enforce the condition $\chi^2_\mathrm{opt} < 5$ to remove faint low-redshift sources showing $\sim$1--1.5$\sigma$ detections in each ACS band.
As defined in \citet{Bouwens2015_LF}, $\chi^2_\mathrm{opt} = \sum \mathrm{abs}(f)/f\, \left(f/\sigma\right)^2$ where $f$ and $\sigma$ represent, respectively, the measured flux density and its error in a given ACS band, while $\mathrm{abs}(f)$ is the absolute value of the flux density.
Finally, we limit our sample to sources with F200W$<$28 as fainter $z\sim6.5-8$ galaxies generally cannot be reliably identified given the typical depths of the ACS imaging ($m\approx28.2$ 3$\sigma$ in each).

We visually inspect each object satisfying the above selection criteria, removing any impacted by artifacts (e.g. diffraction spikes or `snowballs') or that appear to likely be a spurious detection, often resulting from diffuse emission of nearby bright objects.
While it is expected that the surface density of cool brown dwarfs is low in the EGS field ($\sim$0.1 arcmin$^{-2}$; \citealt{Ryan2016_BDsurfaceDensity}), these objects can yield near-infrared colors mimicking a $z\sim7$ Lyman-alpha break \citep[e.g.][]{Stanway2008_contamination}.
We search for any bright point source objects with colors consistent with a brown dwarf solution, using both the empirical SPEX 0.8--2.5$\mu$m brown dwarf spectral library \citep{Burgasser2014_SPEX}, as well as the Sonora\footnote{\url{https://github.com/aburgasser/splat/tree/master/resources/SpectralModels/sonora18}} suite of 0.4--50$\mu$m model brown dwarf spectra \citep{Marley2021_sonora} as references.
One likely brown dwarf is identified with this procedure (reasonably consistent with the expected EGS surface density from \citealt{Ryan2016_BDsurfaceDensity}), and we remove this object from our sample.

We select $z\sim6.5-8$ galaxies from both our standard photometry catalog, as well as that generated after applying our neighbor subtraction algorithm (\S\ref{sec:photometry}).
To ensure that the measured fluxes and colors are reliable, we carefully inspect the postage stamps in every band to ensure that the Kron apertures are not being contaminated by flux from neighboring sources.
We only use photometry from the neighbor-subtracted catalog for objects where, upon visual inspection, it is plausible that flux from neighboring objects could be contaminating the apertures.
For each such object, we also verify that our neighbor-subtraction algorithm is yielding smooth residuals around the Kron apertures (i.e. that it is not significantly over/under-subtracting flux from neighboring sources within or near the apertures).
A small percentage of our remaining candidates fail these checks and are removed from the sample.

Due to the powerful sensitivity and angular resolution of the NIRCam mosaics, a subset of our identified $z\sim6.5-8$ galaxy candidates clearly show multiple emission components separated by $\lesssim$0.5\arcsec{} which are identified as separate systems by \textsc{Source Extractor}.
We treat any such multi-component galaxy as a single system by computing their photometry in manually constructed elliptical apertures that contain the emission from all individual components satisfying our selection criteria described above.
If necessary, we subtract the surface brightness profiles of other neighboring systems if their flux appears to significantly contaminate the manually-constructed apertures.

\begin{figure}
\includegraphics{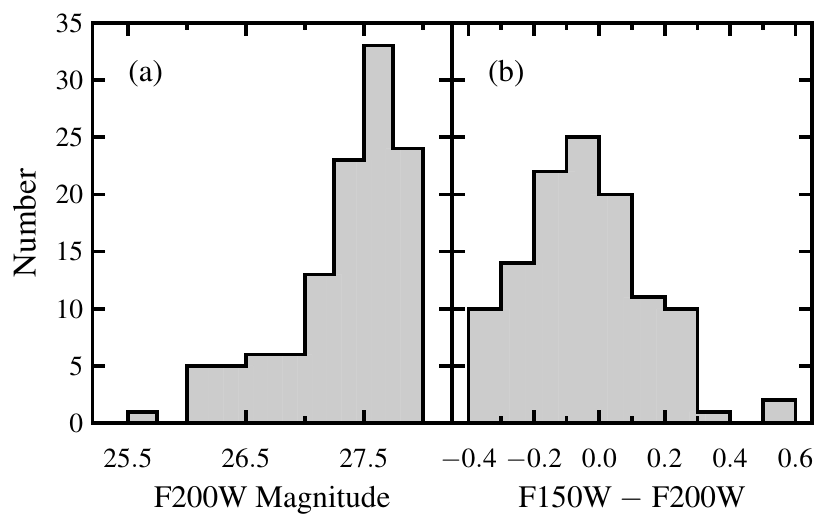}
\caption{The distribution of photometric properties among our 116 Lyman-break $z\sim6.5-8$ galaxies selected with the CEERS/NIRCam data. \textbf{(a)} The distribution of apparent F200W magnitudes (median $m=27.5$). \textbf{(b)} The distribution of F150W$-$F200W colors (median = $-$0.1) which reflect the range of UV slopes among our sample.}
\label{fig:photDistns}
\end{figure}

Our final sample consists of 116 $z\sim6.5-8$ Lyman-break candidates across the June 2022 CEERS footprint. 
Here, we briefly summarize the photometric properties of this sample before using the full multi-band SEDs to infer physical properties in the following sub-section.
The F200W magnitudes of the $z\sim6.5-8$ galaxies range between $m=25.5$ to $m=28.0$ (Fig. \ref{fig:photDistns}a), indicating that our sample spans a factor of $\approx$10 in UV luminosity.
As expected from the steep faint-end slope of the $z\sim7-8$ UV luminosity function \citep[e.g.][]{Finkelstein2015_LF,Ishigaki2018,Bouwens2021_LF}, the sample is heavily weighted towards fainter objects yielding a median F200W $=$ 27.3.
The decline in selection efficiency at $m\gtrsim27.5$ (Fig. \ref{fig:photDistns}a) is expected given the depths of the ACS imaging and our Lyman-break cuts.
The sample also shows a wide range of F150W$-$F200W colors ($-0.40$ to 0.60; Fig. \ref{fig:photDistns}b) which reflects the distribution of rest-UV slopes across our 116 galaxies.
The median F150W$-$F200W color of our $z\sim6.5-8$ sample is $-0.1$, consistent with a nearly flat rest-UV slope ($\beta \approx -2$ where $F_\lambda \propto \lambda^\beta$) as expected from WFC3 analyses of similarly faint systems \citep[e.g.][]{Bouwens2014_beta,Bhatawdekar2021}.

\begin{figure}
\includegraphics{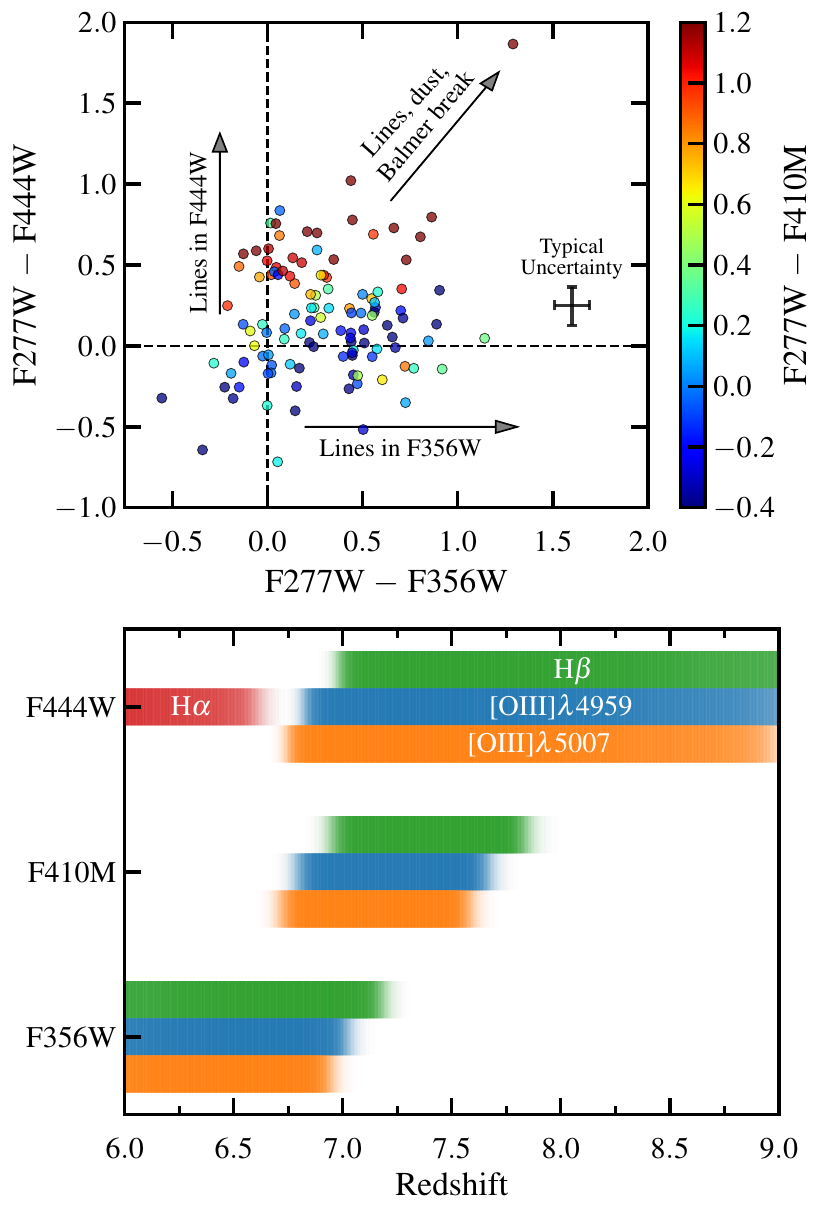}
\caption{\textbf{Top:} A long-wavelength color-color diagram of our $z\sim6.5-8$ CEERS galaxies. The F410M data helps break degeneracies between nebular line emission vs. e.g. dust and Balmer breaks in causing red colors. Dashed lines mark flat colors along each axis. \textbf{Bottom:} Illustration of how the strong optical nebular emission lines H$\alpha$ (red), H$\beta$ (green), [OIII]$\lambda$4959 (blue), and [OIII]$\lambda$5007 (orange) contaminate different NIRCam LW bands at different redshifts. The transparency of the shaded regions vary according to the filter transmission at the observed wavelength of each line.}
\label{fig:filterContamination}
\end{figure}

At $z\sim6.5-8$ the F356W, F410M, and F444W bands fully probe the rest-optical and can be contaminated by strong line emission (i.e. [OIII]$\lambda\lambda$4959,5007, H$\beta$ and H$\alpha$; see Fig. \ref{fig:filterContamination}).
Previous work with \Spitzer{}/IRAC imaging has demonstrated that a significant fraction of the $z\sim7-8$ galaxy population exhibits very high EW line emission ($\gtrsim$1000 \AA{} rest-frame; \citealt{Smit2014,Smit2015,RobertsBorsani2016,Endsley2021_OIII}), which can result in substantial photometric excesses ($\gtrsim$0.5 mag) in the above NIRCam bands, particularly in F410M given its smaller bandwidth.
The impact of such strong line emission is apparent in the F277W$-$F356W vs. F277W$-$F444W color-color space our sample (Fig. \ref{fig:filterContamination}).
We identify objects with very red ($\approx$1) F277W$-$F356W colors yet fairly flat ($\approx$0) F277W$-$F444W colors, likely indicating galaxies at $z\lesssim6.75$ where the combined EW of \OIIIHb{} is boosting F356W considerably more than H$\alpha$ in F444W.
Conversely, a subset of $z\sim6.5-8$ galaxies show very red F277W$-$F444W colors yet flat F277W$-$F356W which are likely objects at $z\gtrsim7.0$ where \OIIIHb{} redshifts out of F356W and into F444W (Fig. \ref{fig:filterContamination}).
There are also objects with red colors in both F277W$-$F356W and F277W$-$F444W which can arise from strong lines at $z\approx6.75-6.90$ where [OIII]$\lambda$5007 falls in both F356W or F444W (Fig. \ref{fig:filterContamination}), though the presence of significant dust attenuation or relatively old stellar populations yielding strong Balmer breaks can lead to such colors as well.
Fortunately, the F410M photometry as well as the rest-UV slopes help differentiate between these various effects as we demonstrate below.

\subsection{Photoionization Modelling} \label{sec:beagle}

We now infer the physical properties of each of the 116 $z\sim6.5-8$ Lyman-break galaxy candidates in our CEERS sample using the SED-fitting code BayEsian Analysis of GaLaxy sEds (\textsc{beagle}; \citealt{Chevallard2016}).
\textsc{beagle} fits the input ACS$+$NIRCam photometry to a suite of SED photoionization models, employing the Bayesian \textsc{multinest} algorithm \citep{Feroz2008,Feroz2009} to produce a posterior probability distribution for each free physical parameter.
The \citet{Gutkin2016} photoionization models used here include both stellar and nebular emission that were computed by processing the latest \citet{BruzualCharlot2003} stellar population synthesis models (see \citealt{Gutkin2016}) through \textsc{cloudy} \citep{Ferland2013}.

Consistent with many previous studies at $z\sim6-8$ \citep[e.g.][]{Ono2012,Labbe2013,Stark2013_NebEmission,Gonzalez2014,Smit2014,Salmon2015,Song2016,deBarros2019,Stefanon2022_sSFR}, we assume a constant star formation history (CSFH) in our \textsc{beagle} fits.
It has been demonstrated that adopting a non-parametric star formation history (SFH) can yield considerably larger stellar masses ($\sim$0.5--1 dex), particularly for objects with very young CSFH ages ($\lesssim$10 Myr) where light from a recent burst is dominating the observed emission \citep{Carnall2019_SFH,Leja2019,Lower2020,Johnson2021,Tacchella2022_SFHs,Topping2022_REBELS,Whitler2023_z10,Whitler2023_z7}.
For these galaxies with very young SEDs, non-parametric SFH fits that disfavor extremely rapid changes in SFR (i.e. the `continuity' prior) tend to add much more star formation at earlier times that is outshined by the recent burst, resulting in stellar masses up to $\approx$2 dex higher than that inferred from CSFH fits.
Therefore, non-parametric SFH models with a continuity prior effectively place an upper limit on the stellar mass permitted by the SED (depending on the assumed formation redshift), whereas CSFH fits place an approximate lower limit.
While dynamical mass and rest-frame near-infrared SED constraints can help distinguish between these two SFHs \citep[e.g.][]{Tang2022,Topping2022_REBELS}, such information does not yet exist for our sample.
Because one of our key goals is to compare the stellar populations dominating the observed rest-UV$+$optical SEDs among our galaxies versus that in brighter $z\sim7-8$ galaxies (\S\ref{sec:sSFRandEW}), we opt to assume a CSFH in our fiducial models.
We will consider the impact of non-parametric SFH models for a subset of our objects in \S\ref{sec:dusty} and \S\ref{sec:stellarMasses}, but note that none of our conclusions rely on absolute measures of stellar mass (or sSFR).

For our \textsc{beagle} fits, we assume a \citet{Chabrier2003} stellar initial mass function with mass range 0.1--300 \Msol{}, a fixed dust-to-metal mass ratio of $\xi_d = 0.3$, and an SMC dust attenuation curve \citep{Pei1992}. 
In \textsc{beagle} the dust opacity is parameterized as the V-band optical depth, $\tau_\mathrm{V}$, which is allowed to vary between 0.001 to 5 with a log-uniform prior.
For ease of understanding what these dust optical depths imply for UV emission, we often quote the attenuation at 1500 \AA{} rest-frame, $A_{1500}$, which is equivalent to 5.2$\tau_\mathrm{V}$ in the SMC attenuation law.
The redshift is allowed to vary in the range $z=5-10$ with a uniform prior, and IGM absorption is incorporated using the empirical model of \citet{Inoue2014}.
We adopt log-uniform priors on both stellar mass and ionization parameter, allowing for values in the range $5 \leq \logMstar \leq 12$ and $-4 \leq \mathrm{log}\ U \leq -1$, respectively.
Galaxy ages are allowed to take values between 1 Myr and the age of the Universe at the sampled redshift with a log-uniform prior as well.
Finally, we adopt a log-uniform prior on metallicity allowing for values of $-2.2 \leq \mathrm{log}(Z/Z_\odot) \leq -0.3$, where we set an upper limit of $\approx$50\% $Z_\odot$ to avoid unreasonable solutions of near-solar metallicity in faint reionization-era galaxies.
In these fiducial \textsc{beagle} fits, the stellar and ISM metallicities are equivalent.

The resulting fiducial \textsc{beagle} fits are generally able to provide a good match to the measured photometry of our Lyman-break $z\sim6.5-8$ galaxies, with a median best-fitting $\chi^2$ of 6.5 (from 10 fitted photometric points) across the full sample and $\chi^2<20$ for 92\% of the objects.
In \S\ref{sec:stellarMasses}, we discuss in detail one object for which our fiducial \textsc{beagle} fits struggle to precisely match the NIRCam photometry and consider other model solutions.
When reporting the inferred physical properties for a given galaxy, we adopt the median of the posterior probability distribution as the fiducial value with $\pm$1$\sigma$ errors taken as the inner 68\% credible interval from the posterior.
To calculate the rest-UV slopes of our galaxies, we fit $F_\lambda \propto \lambda^\beta$ to the photometry of NIRCam bands determined to fall entirely between the \Lya{} line and the Balmer break (i.e. 1216--3600 \AA{} rest-frame) given the inferred redshift specific to each galaxy.
Errors on the rest-UV slope are computed by randomly sampling the redshift from the \textsc{beagle} posterior as well as photometric flux densities given the measured values and their errors.
The absolute UV magnitudes, \Muv{}, are defined at 1500 \AA{} rest-frame and are computed from the output redshift and SED posteriors.
We list a subset of measured and inferred physical properties of all 116 $z\sim6.5-8$ galaxies in our sample in Table \ref{tab:properties}.

\begin{table*}
\centering
\caption{Measured and inferred properties of the 116 $z\sim6.5-8$ CEERS galaxies utilized in this work. The full table is available online.}
\begin{tabular}{P{0.6cm}P{1.3cm}P{1.3cm}P{1.0cm}P{1.0cm}P{1.1cm}P{1.2cm}P{1.5cm}P{1.1cm}P{1.9cm}P{1.1cm}} 
\hline
ID & RA & Dec & F200W & F444W & $z_\mathrm{phot}$ & \Muv{} & log($M_\ast$/$M_\odot$) & sSFR & \OIIIHb{} EW & $A_{1500}$ \Tstrut{} \\
 & [deg] & [deg] & [nJy] & [nJy] & & & & [Gyr$^{-1}$] & [\AA{}] & [mag] \Bstrut{}\\
\hline 
\\[-3pt]
30 & 214.8311941 & 52.7908058 & 37.0$\pm$2.2 & 32.0$\pm$8.4 & 6.57$^{+0.06}_{-0.14}$ & $-19.3^{+0.1}_{-0.1}$ & 7.2$^{+0.1}_{-0.1}$ & 641$^{+213}_{-158}$ & 2526$^{+406}_{-400}$ & 0.1$^{+0.2}_{-0.1}$ \\[5pt]
271 & 214.8248633 & 52.7968353 & 28.0$\pm$2.0 & 25.2$\pm$4.1 & 6.64$^{+0.05}_{-0.06}$ & $-19.4^{+0.0}_{-0.0}$ & 7.9$^{+0.2}_{-0.2}$ & 24$^{+16}_{-11}$ & 547$^{+104}_{-93}$ & 0.0$^{+0.0}_{-0.0}$ \\[5pt]
480 & 214.8116349 & 52.7996698 & 27.8$\pm$2.6 & 70.7$\pm$4.6 & 6.33$^{+0.11}_{-0.09}$ & $-19.3^{+0.1}_{-0.1}$ & 8.5$^{+0.3}_{-1.3}$ & 6$^{+255}_{-4}$ & 300$^{+141}_{-150}$ & 0.1$^{+0.2}_{-0.0}$ \\[5pt]
1091 & 214.8022041 & 52.8058379 & 40.8$\pm$3.2 & 34.3$\pm$4.5 & 6.65$^{+0.03}_{-0.04}$ & $-19.7^{+0.1}_{-0.1}$ & 7.8$^{+0.3}_{-0.2}$ & 42$^{+41}_{-21}$ & 753$^{+135}_{-159}$ & 0.0$^{+0.1}_{-0.0}$ \\[5pt]
1137 & 214.8159100 & 52.8064007 & 227.2$\pm$3.0 & 477.1$\pm$7.0 & 6.96$^{+0.02}_{-0.03}$ & $-21.0^{+0.0}_{-0.0}$ & 10.0$^{+0.0}_{-0.1}$ & 2$^{+0}_{-0}$ & 157$^{+21}_{-12}$ & 1.1$^{+0.1}_{-0.1}$ \\[5pt]
\hline
\end{tabular}
\label{tab:properties}
\end{table*}

\section{Results} \label{sec:results}

\begin{figure}
\includegraphics{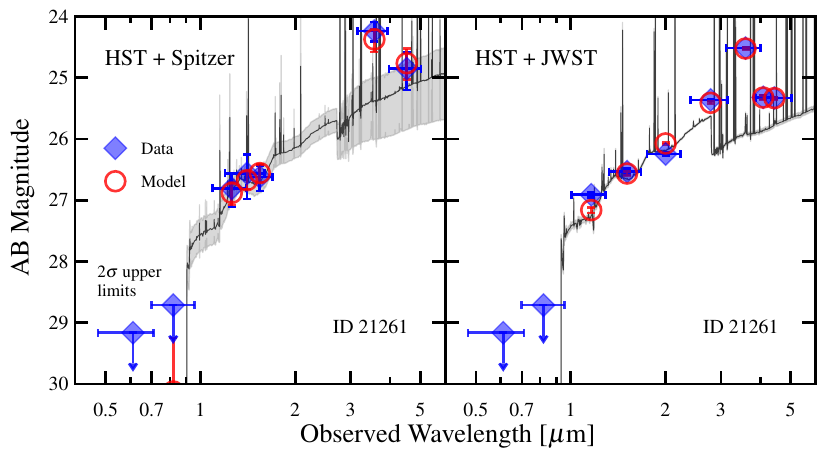}
\includegraphics{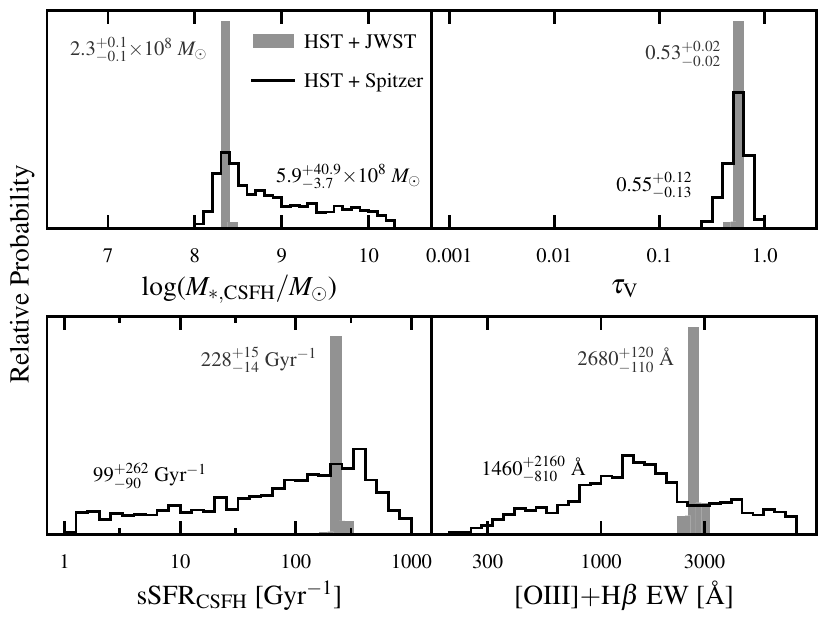}
\caption{Demonstration of how the arrival of \JWST{}/NIRCam data dramatically improves our understanding of the rest-UV$+$optical SEDs and physical properties of typical ($<L_\mathrm{UV}^\ast$) galaxies in the reionization era. In the top panel, we show the previous SED constraints from \HST{}+\Spitzer{} data \citep{Stefanon2017_EGS} (left) versus with new NIRCam photometry (right) for one of the $z\sim6.5-8$ galaxies in our CEERS sample.
The photometric measurements are shown in blue (with 2$\sigma$ upper limits shown in cases of non-detections) while the fitted model photometry and SED from \textsc{beagle} CSFH fits are shown in red and black, respectively. The shaded gray regions and errorbars on the red markers reflect the inner 68\% credible interval from the posterior probability distribution output by \textsc{beagle}. The middle and bottom panels demonstrate how the posteriors on various inferred physical properties tighten substantially with the NIRCam data, illustrating how uncertainties typically decrease from $\approx \pm$0.5--1 dex with \HST{}$+$\Spitzer{} to $\approx \pm$0.05--0.1 dex with ACS+NIRCam.}
\label{fig:HSTSpitzerComparison}
\end{figure}

Due to the limitations of previous near-infrared facilities, it has long been very challenging to constrain the SEDs and hence physical properties (e.g. stellar mass, sSFR, age, nebular line EWs) of typical ($<\!L_\mathrm{UV}^\ast$) galaxies in the reionization era.
\Spitzer{}/IRAC imaging suffered from inadequate sensitivity, strong source confusion, as well as poor sampling of the rest-optical emission with only two broadband filters at 3--5$\mu$m.
While \HST{}/WFC3 was able to deliver relatively sensitive photometry at 1.0--1.6$\mu$m, imaging in moderate-sized ($\sim$100 arcmin$^2$) fields like EGS often reached only $m\sim26.5$ \citep{Grogin2011,Koekemoer2011}, thus limiting insight into the rest-UV emission from typical reionization-era galaxies in these fields as well.
With the tremendous advancements in 1--5$\mu$m imaging sensitivity, angular resolution, and filter suite enabled by \JWST{}/NIRCam, we now begin characterizing the rest-UV+optical SEDs of typical reionization-era galaxies in detail.

Before describing the results of our full $z\sim6.5-8$ CEERS sample in this section, here we first briefly demonstrate how the NIRCam data from CEERS greatly improve our understanding of $<L_\mathrm{UV}^\ast$ $z\sim7-8$ galaxies in EGS relative to what was previously possible with \HST{} and \Spitzer{} data.
In Fig. \ref{fig:HSTSpitzerComparison}, we directly compare the rest UV+optical SED constraints for a typical (F150W$\approx$27) galaxy in our sample when using \HST{} and \Spitzer{} photometry \citep{Stefanon2017_EGS} versus \JWST{}/NIRCam.
Even with the deep \HST{} and \Spitzer{} data over the EGS field, this galaxy is only marginally detected ($\sim$3$\sigma$) in the WFC3 and IRAC 4.5$\mu$m bands, resulting in little constraining power on the shape of the SED.
Consequently, the stellar mass, dust optical depth, sSFR, and \OIIIHb{} EW are all highly uncertain from previous data (see Fig. \ref{fig:HSTSpitzerComparison}).
With the CEERS/NIRCam imaging, this scenario changes dramatically.
The galaxy is detected at S/N$>$20 in every NIRCam band, revealing a strong and highly-significant F356W excess indicative of high-EW \OIIIHb{} emission at $z\approx6.7$, as well as now a clearly red rest-UV slope over the large dynamic range in wavelength space covered by NIRCam, thus confidently implying strong dust attenuation.
From the sensitive 7-band 1--5$\mu$m NIRCam photometry, uncertainties on inferred physical properties decrease to approximately $\pm$0.05--0.1 dex, a vast improvement relative to the typical $\approx$ $\pm$0.5--1 dex uncertainties from \HST{}+\Spitzer{} data (see Fig. \ref{fig:HSTSpitzerComparison}).

In the remainder of this section, we utilize the advanced 1--5$\mu$m imaging capabilities of NIRCam to first characterize the physical properties (e.g. age, stellar mass, dust optical depth) of our large (N=116) $z\sim6.5-8$ CEERS galaxy sample (\S\ref{sec:SEDs}).
In each of the following two sub-sections, we discuss an unexpected population of galaxies identified in our sample: sources with young SEDs and weak \OIIIHb{} emission (\S\ref{sec:weakOIII}) as well as very red galaxies that likely lie at $z\sim7$ given their signatures of strong optical line emission (\S\ref{sec:dusty}).
We then combine the results from our CEERS sample with past studies of the UV-bright $z\sim7-8$ galaxy population to assess whether reionization-era galaxy sSFRs or \OIIIHb{} EWs correlate strongly with UV luminosity (\S\ref{sec:sSFRandEW}).
Finally, we discuss whether we find any evidence of extremely massive ($M_\ast \sim 10^{11}\ M_\odot$) galaxies at $z\sim7-8$, comparing with previous works (\S\ref{sec:stellarMasses}).

\begin{figure*}
\includegraphics{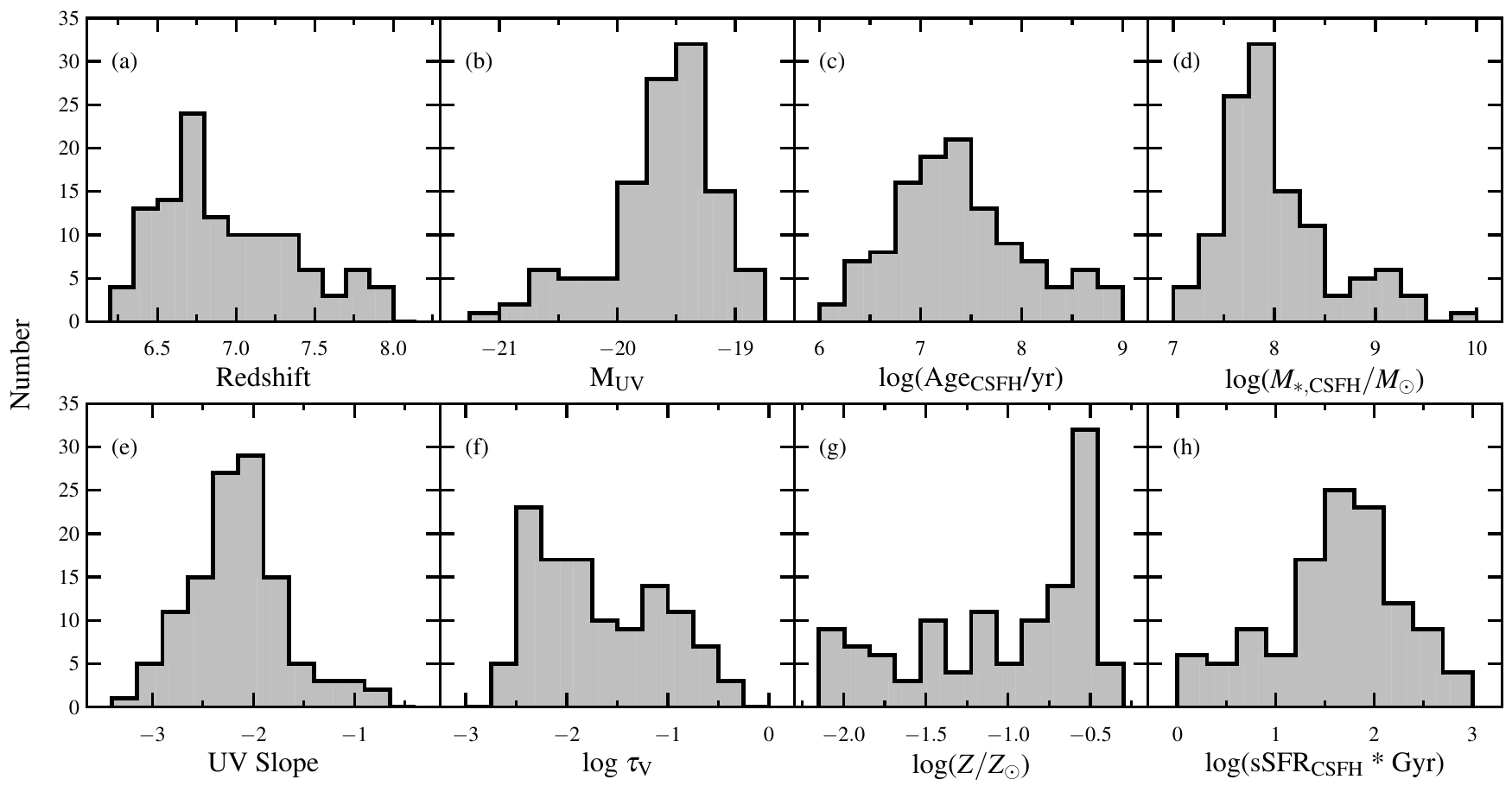}
\caption{Distributions of inferred properties among our sample of 116 Lyman-break $z\sim6.5-8$ galaxies. The values for each galaxy are taken as the median of the posterior probability distribution output by \textsc{beagle} using our fiducial constant star formation history (CSFH) fits (see \S\ref{sec:beagle}).}
\label{fig:inferredPropertyHistograms}
\end{figure*}

\subsection{Demographics of UV-faint \texorpdfstring{$\mathbf{z\sim6.5-8}$}{z ~ 6.5 - 8} Galaxies} \label{sec:SEDs}

\begin{figure*}
\includegraphics{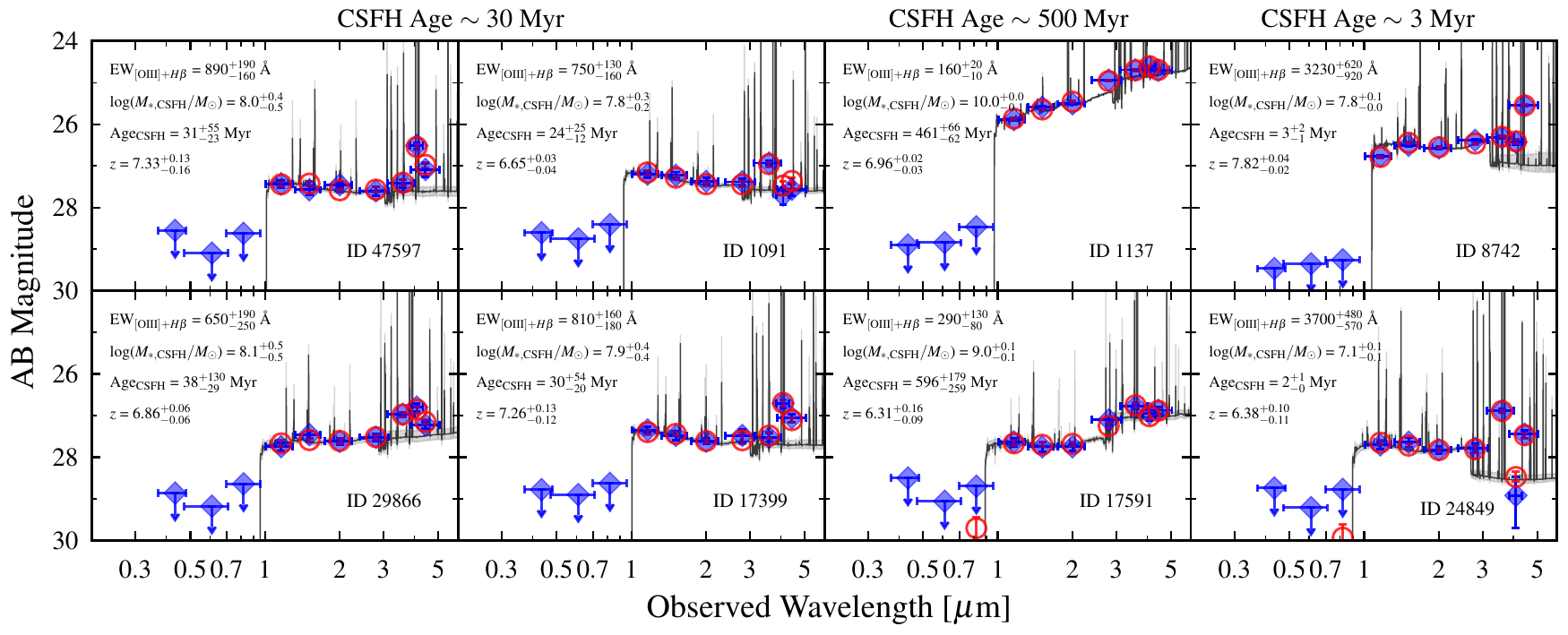}
\caption{Illustration of how the CEERS/NIRCam data reveal a variety of CSFH ages among our $z\sim6.5-8$ CEERS sample. In the left two columns, we show a subset of galaxies with fairly typical CSFH ages ($\sim$30 Myr) among our sample which are evidenced by the lack of a strong Balmer break. A significant subset of our galaxies (29\%) show evidence of prominent Balmer breaks consistent with relatively old CSFH ages (50--550 Myr; see third column). A very similar fraction (28\%) of the galaxies in our sample show long-wavelength photometry suggesting the presence of extremely high-EW nebular line emission consistent with very young CSFH ages ($<$10 Myr; see rightmost column). The SEDs and photometry are shown in the same manner as Fig. \ref{fig:HSTSpitzerComparison}.}
\label{fig:ageSEDs}
\end{figure*}

Before detailing the NIRCam SEDs among our sample of 116 $z\sim6.5-8$ Lyman-break galaxies, we first briefly describe two basic properties of this sample -- their photometric redshifts and absolute UV magnitudes.
As expected from our dropout selection criteria (\S\ref{sec:sample}), the photometric redshifts of galaxies in our sample range between $z=6.27$ and $z=7.97$ (Fig. \ref{fig:inferredPropertyHistograms}a).
The median redshift of the sample is closer to the low end of the distribution ($z=6.82$), consistent with expectations of a declining UV luminosity function at higher redshifts \citep[e.g.][]{Finkelstein2015_LF,Ishigaki2018,Bouwens2021_LF}.
The inferred absolute UV magnitudes of the CEERS $z\sim6.5-8$ galaxies encompass the range $-21.0 \leq \Muv{} \leq -18.8$ with a median $\Muv{} = -19.5$ (Fig. \ref{fig:inferredPropertyHistograms}b), consistent with the F200W magnitude distribution measured among the sample (see Fig. \ref{fig:photDistns}a).
Adopting a characteristic UV luminosity of $L_\mathrm{UV}^\ast = -20.5$ at $z\sim6.5-8$ \citep[e.g.][]{Bowler2017,Bowler2020,Harikane2022_LF}, the typical galaxy in our sample has a UV luminosity of 0.4 $L_\mathrm{UV}^{\ast}$ with the full sample spanning 0.2--1.6 $L_\mathrm{UV}^\ast$.
The vast majority (92\%) of galaxies in our sample are classified as sub-$L_\mathrm{UV}^\ast$ systems.

Now equipped with four deep photometric data points at $\approx$3--5$\mu$m (including in one medium band), we begin investigating the rest-optical SEDs of relatively faint ($\sim$0.4 $L_\mathrm{UV}^{\ast}$) galaxies in the reionization era.
The CEERS/NIRCam data clearly reveal a variety of CSFH ages among UV-faint $z\sim6.5-8$ Lyman-break galaxies (Fig. \ref{fig:ageSEDs}).
For the bulk of our sample, the measured flux density in either F356W, F410M, or F444W is consistent with extrapolating a power-law SED from the rest-UV photometry (see left two columns of Fig. \ref{fig:ageSEDs}).
This implies that young (\ageCSFH{}$\sim$30 Myr) stellar populations dominate their rest-UV+optical SEDs yielding large specific star formation rates (\sSFRCSFH{}$\sim$30 Gyr$^{-1}$).
Nonetheless, we do identify a subset of objects with F356W, F410M, and F444W flux densities boosted by $\sim0.5-1$ mag relative to the rest-UV, with approximately equal excesses in each of the these three bands (see third column of Fig. \ref{fig:ageSEDs}).
Such photometry imply the presence of a prominent Balmer break, consistent with relatively evolved stellar populations (\ageCSFH{}$\sim$100--500 Myr) and low-to-moderate \sSFRCSFH{} ($\sim$2--10 Gyr$^{-1}$).
In addition to sources showing signs of prominent Balmer breaks, we also find a subset of $z\sim6.5-8$ galaxies exhibiting a very strong ($\approx$1--2 mag) photometric excess in F356W, F410M, or F444W, with the extent of the excess differing significantly from band to band (see rightmost column of Fig. \ref{fig:ageSEDs}).
The photometric excess patterns of these galaxies are consistent with contamination from exceptionally high-EW nebular lines (\OIIIHb{} EW$\gtrsim$1500 \AA{}), implying very young CSFH ages ($\lesssim$10 Myr) and correspondingly very large \sSFRCSFH{} ($\gtrsim$100 Gyr$^{-1}$).
We find a very similar fraction of very young (\ageCSFH{}=1--10 Myr) and relatively old (50--550 Myr) galaxies in our sample (28\% and 29\%, respectively), with the remaining 43\% of our sample having moderately young CSFH ages (10--50 Myr; see Fig. \ref{fig:inferredPropertyHistograms}c).

\begin{figure}
\includegraphics{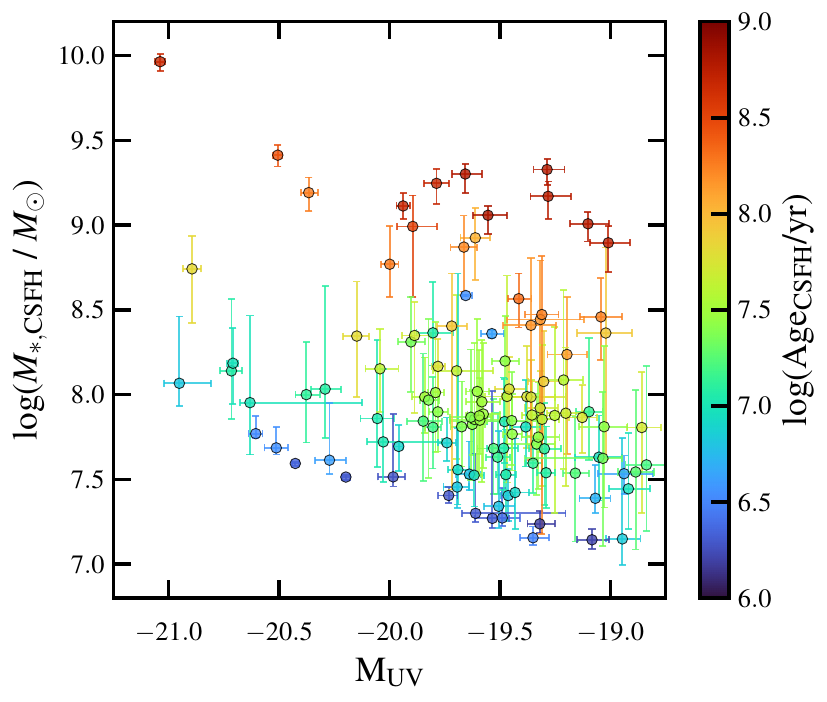}
\caption{The relationship between CSFH stellar mass to UV luminosity among our sample of 116 $z\sim6.5-8$ Lyman-break galaxies in CEERS. Individual points are color-coded by the inferred CSFH age of the corresponding galaxy to illustrate that brighter and/or older galaxies tend to be more massive as expected.}
\label{fig:MstarMuv}
\end{figure}

NIRCam imaging enables not only much stronger constraints on the stellar population ages dominating the SEDs of sub-$L_\mathrm{UV}^\ast$ $z\sim6.5-8$ galaxies, but also on the stellar masses (see Fig. \ref{fig:HSTSpitzerComparison}).
As expected from the typically young CSFH ages ($\sim$30 Myr) and faint UV luminosities ($\Muv{} \sim -19.5$) of our galaxies, we find that our sample is largely composed of low-mass ($M_\ast < 10^9\ M_\odot$) objects in context of our CSFH models.
The median CSFH stellar mass of our galaxies is \MstarCSFH{} = $10^{7.9} \Msol{}$, with individual objects having stellar masses ranging over $>$2 dex from $10^{7.1} \Msol{}$ to $10^{10.0}$ \Msol{} (see Fig. \ref{fig:inferredPropertyHistograms}d).
We show our CEERS galaxies on the stellar mass versus UV luminosity plane in Fig. \ref{fig:MstarMuv}, where we color-code the data point for each galaxy by its CSFH age.
Unsurprisingly, we find that more luminous objects tend to have larger inferred stellar masses.
As discussed in \S\ref{sec:beagle}, for galaxies with very young SEDs our CSFH fits are only modeling light from a recent burst resulting in low-mass ($\lesssim10^8\ M_\odot$) solutions, though it is likely that some of these systems have true stellar masses that are substantially ($\gtrsim$3$\times$) larger.
In sections \S\ref{sec:dusty} and \S\ref{sec:stellarMasses}, we consider the impact of adopting non-parametric SFH models for a subset of our sample.

Because the CEERS/NIRCam data deliver sensitive imaging across 1--5$\mu$m, we can investigate not only the rest-optical SEDs of our galaxies, but their rest-UV slopes as well.
As expected from previous studies using \HST{}/WFC3 data (\citealt{McLure2011,Dunlop2012,Finkelstein2012,Rogers2013,Bouwens2014_beta,Bhatawdekar2021}), we find that our relatively faint ($\Muv{} \sim -19.5$) CEERS galaxies tend to have blue UV slopes ($\beta \sim -2$; see Fig. \ref{fig:inferredPropertyHistograms}e).
Consequently, our \textsc{beagle} SED fitting results imply little dust attenuation in the majority of our sample (median $\tau_\mathrm{V} = 0.013$ or $A_{1500} = 0.07$ mag; Fig. \ref{fig:inferredPropertyHistograms}f).
The $\beta$ vs. \Muv{} relation of our $z\sim6.5-8$ galaxies is presented in \citet{Topping2022_blueSlopes}.
Therein, we discussed a subset of extremely blue objects ($\beta \lesssim -3$) exhibiting NIRCam SEDs consistent with not only negligible dust attenuation, low metallicities ($\lesssim$5\% $Z_\odot$), and young CSFH ages ($\lesssim$20 Myr), but also high ionizing photon escape fractions ($\fesc{} \gtrsim 50$\%) which mitigate reddening from nebular continuum emission.

In each of the following two sub-sections, we discuss an unexpected population of $z\sim6.5-8$ Lyman-break galaxies identified in our sample that we have not yet detailed. 
The first population we discuss are objects with NIRCam SEDs suggesting both young CSFH ages yet relatively weak \OIIIHb{} emission.
Next, we describe a surprisingly large number of galaxies with very red UV slopes, as well as strong long-wavelength photometric excesses which likely place each system at $z\sim7$.
We argue that the abundance of each of these galaxy classes may have important implications for our understanding of star formation, ionizing photon escape, and chemical evolution in the reionization era.

\begin{figure*}
\includegraphics{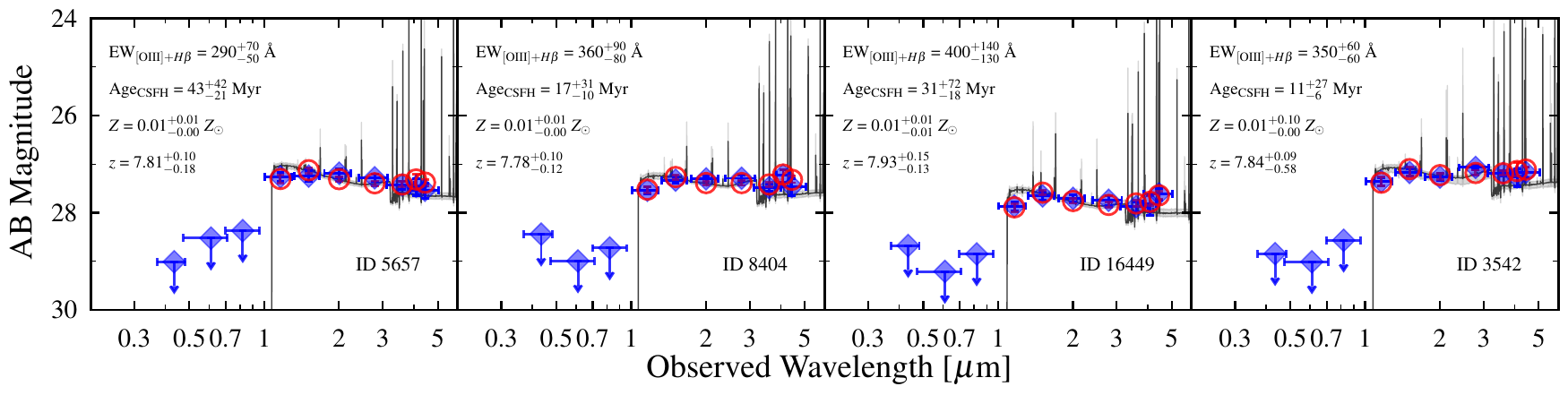}
\caption{Four examples of $z\sim6.5-8$ Lyman-break galaxies with NIRCam SEDs implying both young CSFH ages ($\lesssim$50 Myr) and relatively low \OIIIHb{} EW ($<$600 \AA{}). Such objects comprise 18\% of our total CEERS sample, suggesting that they may be a significant class of sub-$L_\mathrm{UV}^\ast$ reionization-era galaxies, though redshift confirmation is necessary. In the context of our fiducial \textsc{beagle} fits, these SEDs can be reproduced at $z\sim6.5-8$ with extremely low metallicities ($\approx$1--3\% $Z_\odot$; see also Fig. \ref{fig:age_EW_metallicity}). However, there are alternative physical explanations as illustrated in Fig. \ref{fig:youngWeakOIIISEDs_otherExplanations}.}
\label{fig:youngWeakOIIISEDs}
\end{figure*}

\begin{figure}
\includegraphics{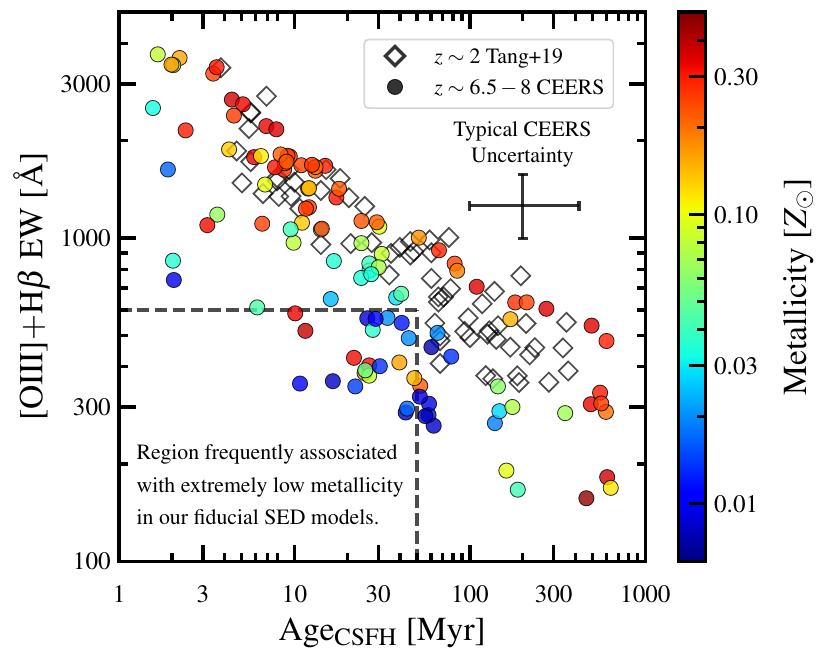}
\caption{The relationship between \OIIIHb{} EW and CSFH age. The $z\sim6.5-8$ galaxies in our UV-selected Lyman-break CEERS sample are shown with filled-in circles, while $z=1.3-2.4$ galaxies from the [OIII]-selected sample of \citet{Tang2019} are shown as open diamonds. A considerable fraction (18\%) of our CEERS sample is composed of galaxies with young SEDs (\ageCSFH{}$<$50 Myr), yet also relatively weak \OIIIHb{} (EW$<$600 \AA{}) while such a population seems comparatively rare at $z\sim2$. In our fiducial \textsc{beagle} fits, such SEDs are frequently reproduced with extremely low metallicities ($\sim$1--3\% $Z_\odot$) as shown by the color coding.}
\label{fig:age_EW_metallicity}
\end{figure} 

\begin{figure}
\includegraphics{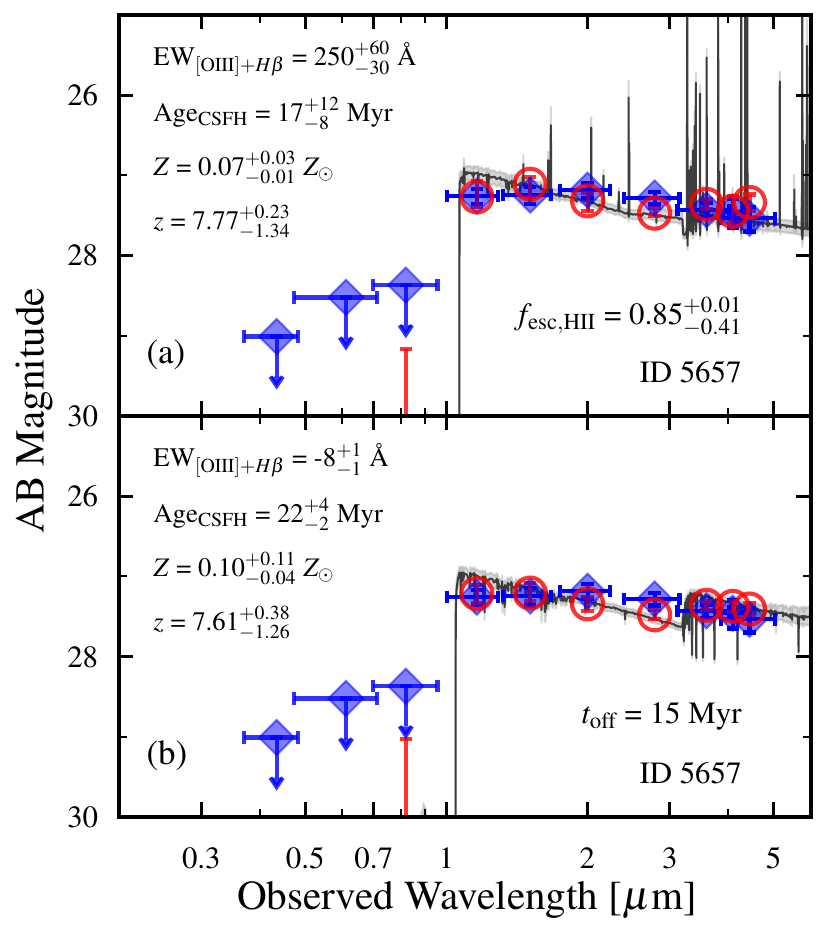}
\caption{We find that extremely low metallicity is not the only plausible explanation for the NIRCam SEDs implying young CSFH ages and weak \OIIIHb{} among our $z\sim6.5-8$ candidates. The top panel demonstrates how the photometry of at least some systems can be reproduced with models allowing for a high fraction of ionizing photons to leak from HII regions within the galaxy ($f_\mathrm{esc,HII} \sim 0.5$; see also \citealt{Topping2022_blueSlopes}). The bottom panel shows the NIRCam photometry for the same object also reproduced by a model where the star formation rate has strongly declined over the past 15 Myr resulting in a deficit of OB stars. Future spectroscopic \JWST{} observations will be able to not only confirm the redshifts of these objects, but also help distinguish between the three possible physical explanations we have proposed for their SEDs.}
\label{fig:youngWeakOIIISEDs_otherExplanations}
\end{figure}

\subsection{The Nature of Weak [OIII]+H\texorpdfstring{$\mathbf{\beta}$}{beta} Emitters at \texorpdfstring{$\mathbf{z\sim6.5-8}$}{z ~ 6.5 - 8}} \label{sec:weakOIII}

Even prior to the launch of \JWST{}, it was relatively easy to identify $z\sim7-8$ galaxies with very high \OIIIHb{} EWs ($>$1000 \AA{}) given that these objects would show a substantial ($\gtrsim$1 mag) excess in one of the \Spitzer{}/IRAC bands, yielding a bright (and relatively high S/N) detection \citep[e.g.][]{Smit2014,Smit2015,RobertsBorsani2016,deBarros2019,Strait2020,Endsley2021_OIII}.
However, it has remained very difficult to quantify the population of sub-$L_\mathrm{UV}^\ast$) reionization-era galaxies with low \OIIIHb{} EWs given that these sources would generally remain poorly detected in IRAC.
With the highly advanced 3--5$\mu$m imaging capabilities of NIRCam, it is now possible to not only easily identify, but also characterize weak \OIIIHb{} emitters among sub-$L_\mathrm{UV}^\ast$ reionization-era galaxies.
As expected, we identify a subset of galaxies in our CEERS $z\sim6.5-8$ sample that exhibit NIRCam SEDs consistent with both relatively old stellar populations (\ageCSFH{}$>$300 Myr) and weak \OIIIHb{} emission (EW$<$500 \AA{}; see e.g. IDs 1137 and 17591 in Fig. \ref{fig:ageSEDs}).
The comparatively weak nebular lines in these evolved objects are expected given that the rest-optical continuum will be boosted over time by the build-up of A stars, while the nebular line luminosity will remain roughly constant as OB stars undergo short life cycles.
However, as detailed below we also find a seemingly significant population of UV-faint $z\sim6.5-8$ Lyman-break galaxies with SEDs implying both young CSFH ages and weak \OIIIHb{}.

The CEERS NIRCam data reveal several candidate $z\sim6.5-8$ galaxies with strong Lyman-alpha breaks (F814W$-$F115W=1.5--2.7; median 1.8) as well as SEDs implying young CSFH ages yet also little-to-no photometric excesses in the long-wavelength bands.
We show the SEDs of four such objects in our sample in Fig. \ref{fig:youngWeakOIIISEDs} where it can be seen that their F356W, F410M, or F444W photometry are reasonably consistent with a power-law SED extrapolated from the shorter-wavelength NIRCam photometry. 
The lack of significant Balmer breaks in these systems indicates that their light-weighted ages are young (\ageCSFH{}$<$50 Myr), implying that a substantial portion of their emergent light should be coming from recently-formed OB stars which are highly efficient producers of LyC photons.
Accordingly, we might expect to see strong nebular line emission in such seemingly young objects, as indeed found among a sample of UV-faint (median $\Muv{} \approx -19.5$) extreme emission line galaxies (EELGs) at $z\sim2$ where \OIIIHb{} EWs$>$800 \AA{} are nearly ubiquitous at \ageCSFH{}$<$50 Myr (\citealt{Tang2019}; see Fig. \ref{fig:age_EW_metallicity}).
However, we find little-to-no significant photometric excesses in the long-wavelength bands for several of the young $z\sim6.5-8$ galaxies in our sample, implying considerably lower \OIIIHb{} EWs at fixed CSFH age relative to $z\sim2$.
From our fiducial \textsc{beagle} fits, young (\ageCSFH{}$<$50 Myr), relatively weak \OIIIHb{} emitters (EW$<$600 \AA{}) comprise 18\% (21/116) of our CEERS sample, suggesting that they may be a significant class of sub-$L_\mathrm{UV}^\ast$ reionization-era galaxies that are comparatively rare at $z\sim2$.
While the \citet{Tang2019} $z\sim2$ EELG sample mentioned above was selected on [OIII] emission, the adopted selection threshold of [OIII]$\lambda$5007 EW$>$225 \AA{} still allows for the inclusion of relatively weak \OIIIHb{} emitters (EW$\approx$350--600 \AA{}) and such galaxies were indeed found with \ageCSFH{}$\sim$100--500 Myr (Fig. \ref{fig:age_EW_metallicity}).
Below, we discuss potential physical origins for the weak \OIIIHb{} and young SEDs that appear present in our $z\sim6.5-8$ sample, though we emphasize that spectroscopic follow-up will be necessary to confirm their redshifts.

In the context of our fiducial \textsc{beagle} fits, the combination of young CSFH age ($<$50 Myr) and relatively low \OIIIHb{} EW ($<$600 \AA{}) is frequently reproduced with extremely low metallicities ($\approx$1--3\% $Z_\odot$; see Fig. \ref{fig:age_EW_metallicity}).
In such metal-poor models, [OIII] emission is greatly weakened relative to the more chemically evolved ($\sim$20\% $Z_\odot$) systems with similarly high \sSFRCSFH{} ($\gtrsim$20 Gyr$^{-1}$) found at both lower redshifts ($z\sim2$) and bright UV magnitudes ($\Muv{} \lesssim -21$) at $z\sim7-8$ (e.g. \citealt{Endsley2021_OIII,Tang2021_UVlines}).
Even though H$\beta$ EWs increase with decreasing metallicity, the relative contribution of [OIII]$\lambda\lambda$4959,5007 to \OIIIHb{} is generally far more dominant ([OIII]/H$\beta$ $\sim$ 5--10; e.g. \citealt{Steidel2016,Maseda2018,Sanders2018,Tang2019}) such that extremely low metallicities result in much smaller total \OIIIHb{} EWs at fixed \sSFRCSFH{}.

While our fiducial \textsc{beagle} models demonstrate that extremely low metallicity solutions can reproduce relatively weak \OIIIHb{} emission at young CSFH ages (see Figs. \ref{fig:youngWeakOIIISEDs} and \ref{fig:age_EW_metallicity}), here we discuss two alternative physical explanations.
Because nebular emission lines are byproducts of the interaction between hydrogen ionizing photons and gas in the ISM, these lines can be weakened at high LyC escape fractions (\fesc{}; e.g. \citealt{Zackrisson2013}).
If the LyC leakage is driven by density-bounded conditions, we would expect to see weaker [OIII] and H$\beta$ emission at very high \fesc{} \citep{Plat2019}.
In Fig. \ref{fig:youngWeakOIIISEDs_otherExplanations}a, we illustrate how high escape fractions ($\fesc{} \sim 0.5$) can also plausibly explain weak \OIIIHb{} in a subset of $z\sim6.5-8$ galaxies with young SEDs (see also \citealt{Topping2022_blueSlopes}).
Here, we are using the \textsc{beagle} photoionization models introduced in \citet{Plat2019} which allow for non-zero ionizing photon escape from HII regions within the galaxy (i.e. $f_\mathrm{esc,HII} > 0$) due to density-bounded conditions.
While such ionizing photons must also travel through the ISM and CGM to reach the intergalactic medium, sources with large $f_\mathrm{esc,HII}$ would be prime candidates for key contributors to reionization. 

Another way nebular line emission can be weakened in a seemingly young galaxy is if that object has experienced a sharp decline in star formation rate within the past $\sim$10 Myr, resulting in very few (if any) remaining OB stars.
To illustrate how such a star formation history could reproduce young CSFH SEDs with weak \OIIIHb{}, we take our fiducial \textsc{beagle} fitting setup but force the SFR to 0 $M_\odot$/yr over the most recent 15 Myr (with a constant non-zero SFR at $>$15 Myr).
Such a recent turnoff in star formation can produce SEDs that provide a satisfactory match the observed NIRCam photometry for a subset of our young weak \OIIIHb{} emitters (see Fig. \ref{fig:youngWeakOIIISEDs_otherExplanations}b).
In these solutions, the observed SED is dominated by light from a population of stars that started forming $\sim$50 Myr ago and assembled $\sim$10$^9$ $M_\odot$ before star formation plummeted in the most recent 15 Myr.
From an empirical standpoint, such bursty star formation histories are consistent with the presence of extremely high \sSFRCSFH{} ($>$100 Gyr$^{-1}$) systems we identify in our sample.
With implied mass-doubling times of $<$10 Myr, these galaxies are caught in a phase of intense star formation activity that likely was proceeded and will be followed by periods with much lower SFRs.
A subset of our young, weak \OIIIHb{} emitters may reflect these periods of relatively inactive star formation.
Consistent with this picture, recent high-resolution hydrodynamic simulations also predict that $z\sim7-8$ galaxies frequently undergo episodes of extreme star formation between periods of weak-to-moderate star formation \citep[e.g.][]{Ceverino2018,Ma2018_burstySFH}. 

It is currently not clear which of these three physical scenarios is responsible for the apparent young SEDs with weak \OIIIHb{}.
In principle, low metallicities, high \fesc{}, and bursty star formation histories could contribute to the effect in different galaxies.
Fortunately, future \JWST{} surveys will soon allow us to not only confirm the existence of these reionization-era systems, but also better understand the physical origin of their SEDs.
For example, direct temperature measurements could test for the presence of extremely metal-poor populations \citep{ArellanoCordova2022,Schaerer2022,Taylor2022,Curti2023,Rhoads2023,Trump2023}. 
Moreover, with very deep NIRSpec spectra it will be possible to determine if O stars are contributing significantly to the rest-UV continuum emission by searching for P-Cygni profiles of e.g. CIV$\lambda$1540.
A complete lack of such O-star features would be consistent with the bursty star formation history scenario discussed above, while we would expect O stars to contribute significantly to the rest-UV continuum in the very high \fesc{} and extremely low metallicity scenarios.

\begin{figure}
\includegraphics{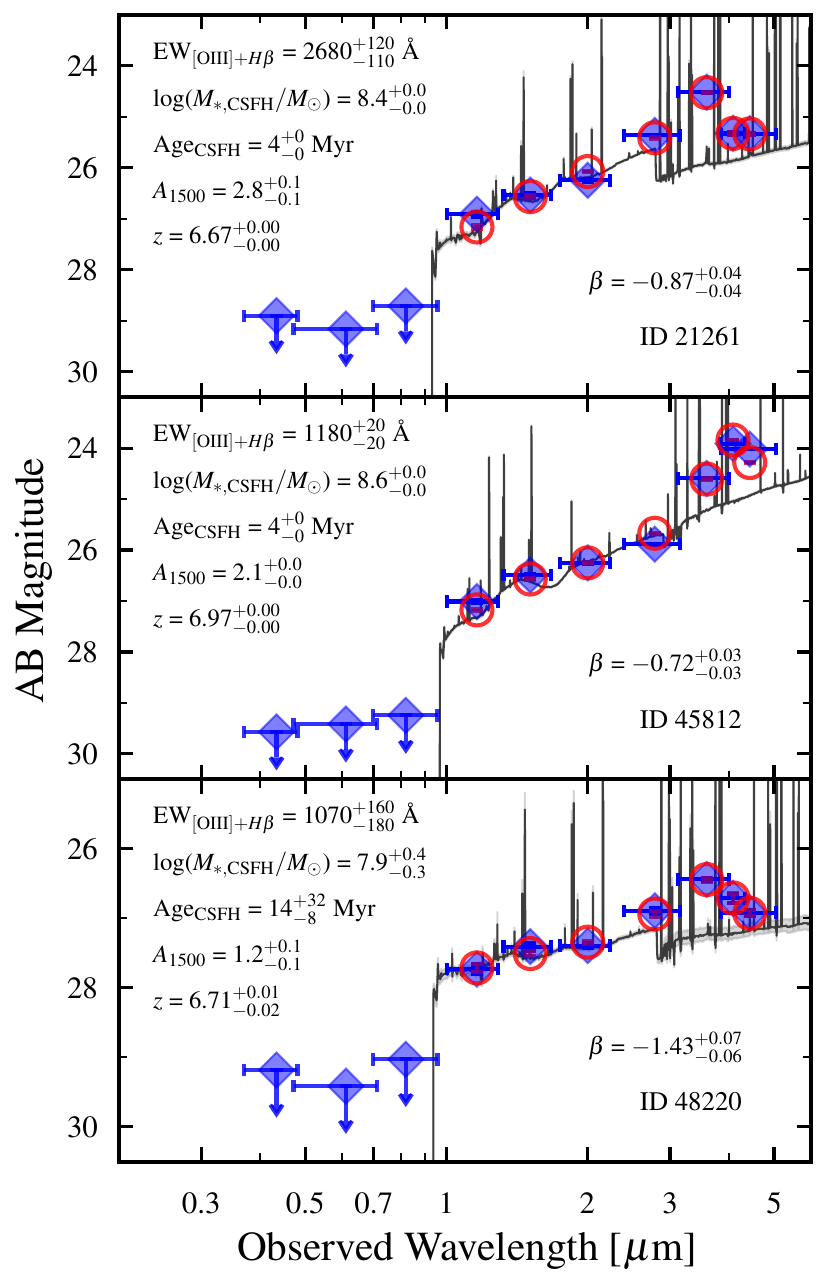}
\caption{The three $z\sim6.5-8$ Lyman-break galaxies in our CEERS sample with both very red UV slopes ($-1.4 \leq \beta \leq -0.7$) and strong \OIIIHb{} emission (EW$>$800 \AA{}). The strong dust reddening inferred by our \textsc{beagle} fits imply that $\approx$70--90\% of their star formation is obscured. Even though these UV-faint ($\approx$0.5 $L_\mathrm{UV}^\ast$) objects represent only 3\% of our CEERS sample by number, their total obscured SFR makes up 19\% of all (unobscured$+$obscured) SFR across our full sample, suggesting they are important contributors to the star formation rate density at early times.}
\label{fig:redEELGseds}
\end{figure}

\subsection{Very Red UV-faint \texorpdfstring{$\mathbf{z\sim6.5-8}$}{z ~ 6.5-8} Galaxies in Small Areas} \label{sec:dusty}

We identify three very red $z\sim6.5-8$ galaxies with strong line emission in our CEERS sample.
The NIRCam SEDs of these three objects are shown in Fig. \ref{fig:redEELGseds} where they all exhibit long-wavelength photometric excesses implying \OIIIHb{} EW$>$800 \AA{} as well as high S/N ($\gtrsim20\sigma$) short-wavelength photometry implying red UV slopes ($\beta$ $\geq$ $-1.4$).
The combination of very strong Ly$\alpha$ breaks (F814W$-$F115W $\geq$ 2.1) and clear signatures of strong nebular line contamination in each of these galaxies gives us confidence that they are robust reionization-era systems.
The photometric redshifts and absolute UV magnitudes of these galaxies span $z$ = 6.7--7.0 and $-19.7 \leq \Muv{} \leq -19.1$, respectively, indicating that they all fall under the classification of sub-$L_\mathrm{UV}^\ast$ sources.
However, as expected from their very red UV-slopes ($-1.4 \leq \beta \leq -0.7$) all four systems are inferred to be heavily reddened ($A_{1500}$ = 1.0--2.8 mag) implying that they would have appeared as UV-bright ($-22.0 \lesssim \Muv{} \lesssim -20.5$) galaxies in the absence of dust.

Given the implied large dust reservoirs in these systems, we would perhaps expect them to be fairly massive ($M_\ast\gtrsim$10$^{9}$ $M_\odot$; e.g. \citealt{Dayal2022_REBELS,Ferrara2022_REBELS}).
While our CSFH \textsc{beagle} fits suggest that these objects have relatively low stellar mass ($\approx$1--4$\times$10$^{8}$ $M_\odot$), such fits account for only the recent burst suggesting that significantly higher stellar mass solutions may be feasible \citep[e.g.][]{Carnall2019_SFH,Tacchella2022_SFHs,Whitler2023_z7}. 
Indeed, we find that stellar masses of $\approx$10$^{9}$ $M_\odot$ to $\approx$3$\times$10$^{10}$ $M_\odot$ are possible when adopting \textsc{prospector} non-parametric SFH models utilizing the continuity prior.
For these non-parametric fits, we adopt the setup described in \citet{Whitler2023_z7} where there are eight time bins for the SFH, with the two most recent time bins fixed to 0--3 Myr and 3--10 Myr ago, and the rest evenly spaced in logarithmic lookback time out to an assumed formation redshift of $z_\mathrm{form} = 20$.

In recent years, a growing number of $z\sim7-8$ Lyman-break galaxies have been identified with fairly high S/N ($\gtrsim$5$\sigma$) near-IR photometry implying very red UV slopes ($\beta \sim -1$; e.g. \citealt{Bowler2014,Smit2015,Stefanon2019,Endsley2021_OIII}) and a considerable number of these objects are now spectroscopically confirmed via far-infrared follow-up \citep{Smit2018,Bouwens2022_REBELS,Fujimoto2022,Endsley2023_radioConfirmation}.
However, all of these previously-known very red $z\sim7-8$ systems appear UV luminous ($\Muv{} < -21$) even in the presence of the substantial dust reddening.
This is a direct result of the limited near-IR sensitivity of ground-based cameras as well as \HST{}/WFC3, making it very challenging to confidently identify red $z\sim7-8$ galaxies at more typical UV luminosities ($\Muv{} \gtrsim -20$). 
NIRCam's greatly improved near-IR sensitivity (and SED sampling) relative to WFC3 changes this scenario dramatically as illustrated in Fig. \ref{fig:HSTSpitzerComparison}.
One of our very red CEERS galaxies (ID 21261) is only marginally detected in the CANDELS EGS WFC3 data (S/N$\sim$3) such that it was previously impossible to quantify the UV slope with high confidence.
With CEERS/NIRCam data, this object is detected at S/N$>$20 in F115W, F150W, and F200W, clearly revealing a strong increase in flux density over $\approx$1500--2500 \AA{} rest-frame indicating a very red UV slope ($\beta = -0.87 \pm 0.04$) and strong dust attenuation ($A_{1500} \approx$ 2.8).

The fact that the NIRCam data reveal three very red ($-1.4 \leq \beta \leq -0.7$) $z\sim6.5-8$ galaxies in the CEERS area suggests that there remains a significant tail of heavily dust-reddened galaxies down to at least $\approx$0.5 $L_\mathrm{UV}^\ast$ in the reionization era (see also \citealt{Barrufet2023,Rodighiero2023,PerezGonzalez2023}).
This may have important consequences for our understanding of the extent to which obscured star formation contributes to stellar mass assembly in the very early Universe.
The dust optical depths inferred by our \textsc{beagle} fits imply that $\approx$70--90\% of the star formation in our three young, very red CEERS galaxies is obscured, resulting in individual obscured CSFH SFRs of 4--92 $M_\odot$ yr$^{-1}$ for each object.
Even though these young, very red systems represent only 3\% of our CEERS sample by number, their combined obscured SFR$_\mathrm{CSFH}$ (145 $M_\odot$ yr$^{-1}$) makes up 19\% of all (unobscured$+$obscured) inferred stellar mass growth across our full sample (745 $M_\odot$ yr$^{-1}$).
The obscured SFRs of our young, very red galaxies may even be underestimated from the \textsc{beagle} SED fits.
ALMA follow-up has shown that the total SFR inferred from the rest-UV+optical SED is often a factor of $\approx$2 lower than the SFR derived by combining UV and far-infrared data, at least among very UV-bright ($\Muv{} \sim -22$) $z\sim7-8$ galaxies \citep{Topping2022_REBELS}.
Moreover, at least one very red UV-bright $z\simeq7$ galaxy (COS-87259; $\beta = -0.6 \pm 0.3$) has been found to be extremely bright in the far-infrared with SCUBA-2 and ALMA data, implying an obscured star formation rate of $\approx$1300 $M_\odot$ yr$^{-1}$ (\citealt{Endsley2022_radioAGN,Endsley2023_radioConfirmation}; see also the similar object in GOODS-N described in \citealt{Fujimoto2022}).
Deep far-infrared follow-up will be critical to assess the extent of obscured star formation in the relatively UV-faint ($\lesssim$0.5 $L_\mathrm{UV}^\ast$) and very red $z\sim7$ galaxies identified from deep \JWST{} data.

\subsection{The \texorpdfstring{M$\mathbf{_\mathrm{UV}}$}{Muv} Dependence of sSFRs and [OIII]+H\texorpdfstring{$\mathbf{\beta}$}{beta} EWs} \label{sec:sSFRandEW}

Within the past few years, it has become possible to statistically characterize the sSFRs and \OIIIHb{} EWs of UV-bright ($-23 \lesssim \Muv{} \lesssim -21$) $z\sim7-8$ galaxies thanks to the growing collection of $>$deg$^2$ data sets with deep optical, near-infrared, and IRAC imaging (e.g. \citealt{Mauduit2012,McCracken2012,Jarvis2013,Steinhardt2014,Ashby2018,Aihara2022}).
These UV-luminous reionization-era galaxies tend to exhibit SEDs consistent with high \OIIIHb{} EWs ($\approx$700 \AA{}) as well as young CSFH ages and hence large \sSFRCSFH{} (\citealt{Stefanon2019,Endsley2021_OIII,Topping2022_REBELS,Whitler2023_z7}).
In the more local Universe ($z\lesssim2$), it has been shown that such galaxy properties are clearly associated with high ionizing photon production efficiency \citep{Chevallard2018_z0,Tang2019}, as well as perhaps higher LyC escape fractions \citep[e.g.][]{Izotov2016b,Flury2022_LyCdiagnostics,Saxena2022}.
Unfortunately, our understanding of sSFRs and \OIIIHb{} EWs at fainter magnitudes has been hindered by IRAC's poor sensitivity, leaving much uncertainty on the possible extent to which galaxy ionizing properties correlate with UV luminosity in the reionization era \citep[e.g.][]{Finkelstein2019,Naidu2020}.
Now equipped with deep NIRCam imaging, we can begin constraining the demographics of sSFRs and \OIIIHb{} EWs among sub-$L_\mathrm{UV}^\ast$ $z\sim7-8$ galaxies to compare with the more UV-luminous population.

\begin{figure}
\includegraphics{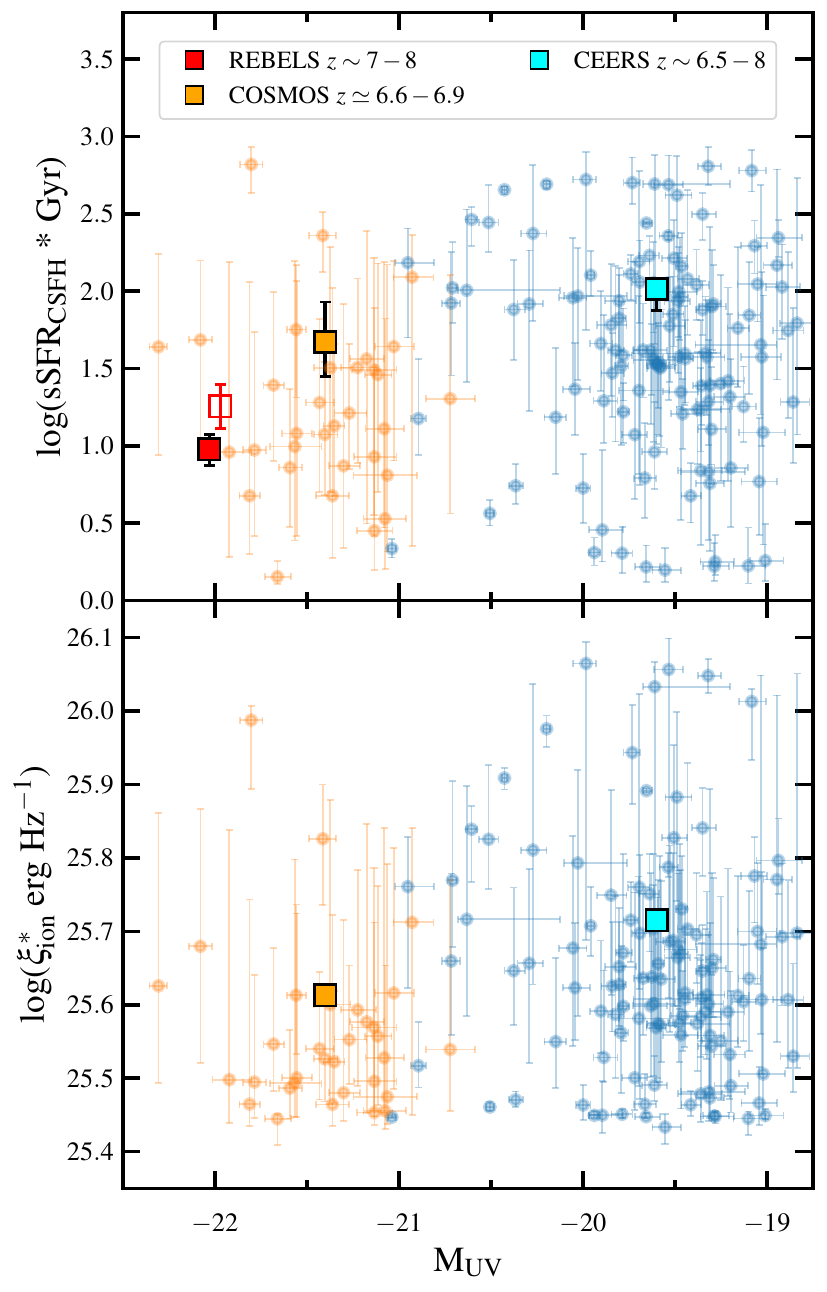}
\caption{\textbf{Top:} The correlation between CSFH sSFR and UV luminosity at $z\sim7-8$ using values inferred from \textsc{beagle} fits to the rest-UV$+$optical SED. We combine results from our CEERS sample with that of the brighter COSMOS \citep{Endsley2021_OIII,Whitler2023_z7} and REBELS \citep{Bouwens2022_REBELS,Topping2022_REBELS} samples, finding evidence of a strong ($\gtrsim$5$\times$) increase in \sSFRCSFH{} between $\sim$4 $L_\mathrm{UV}^\ast$ and $\sim$0.4 $L_\mathrm{UV}^\ast$. The filled-in squares show the median \sSFRCSFH{} of each sample, while we show points for individual galaxies in the CEERS and COSMOS sample. The open red square shows the median \sSFRCSFH{} inferred from the ALMA REBELS sample using unobscured$+$obscured SFRs computed from UV and IR data, and is slightly offset in \Muv{} from the filled red marker for visual clarity. \textbf{Bottom:} The inferred correlation between intrinsic stellar ionizing photon production efficiency (corrected for dust and nebular continuum emission) and UV luminosity at $z\sim7-8$ assuming a CSFH. We find evidence for a modest ($\approx$0.1 dex) increase in the typical \xiion{} value between $\sim$2 $L_\mathrm{UV}^\ast$ and $\sim$0.4 $L_\mathrm{UV}^\ast$.}
\label{fig:sSFR}
\end{figure}

We find that the NIRCam SEDs of CEERS $z\sim6.5-8$ galaxies generally imply very large CSFH specific star formation rates, with 63\% of our sample having a fiducial \sSFRCSFH{}$>$30 Gyr$^{-1}$ (see Fig. \ref{fig:inferredPropertyHistograms}h).
Following previous studies \citep{Topping2022_REBELS}, we compute the median \sSFRCSFH{} of our sample and its associated uncertainty by bootstrap resampling values from the \textsc{beagle} posteriors.
That is, we randomly select 116 galaxies with replacement from our CEERS sample, pulling a \sSFRCSFH{} from the \textsc{beagle} posterior probability distribution for each galaxy, and repeat this process 10000 times.
This yields a median and inner 68\% credible interval $\sSFRCSFH{} = 103^{+9}_{-28}$ Gyr$^{-1}$ (Fig. \ref{fig:sSFR}), consistent with a scenario in which the bulk of UV-faint ($\sim$0.4 $L_\mathrm{UV}^\ast$) reionization-era systems have recently experienced a strong upturn in their star formation activity.

We compare the \sSFRCSFH{} derived among our relatively faint CEERS sample (median $\Muv{} = -19.5$) with results obtained from very UV-luminous $z\sim7$ galaxies whose rest-optical properties have been characterized with IRAC data.
Leveraging ultra-deep optical narrow-band imaging across the 1.5 deg$^2$ COSMOS UltraVISTA field, we previously assembled a large sample of UV-bright ($-22.5 \lesssim \Muv{} \lesssim -21$) Lyman-break galaxies at $z\simeq6.6-6.9$ where the IRAC 3.6$\mu$m band is contaminated by \OIIIHb{} while the 4.5$\mu$m band cleanly probes the rest-optical continuum \citep{Endsley2021_OIII}.
After only considering objects with robust IRAC deconfusion, we obtain a final sample of 36 UV-bright (median $\Muv{} = 21.4$) $z\simeq6.6-6.9$ COSMOS galaxies (see \citealt{Whitler2023_z7}).
We fit these COSMOS galaxies with \textsc{beagle} adopting the same prior set that we use for the CEERS sample, finding a median inferred $\sSFRCSFH{} = 47^{+38}_{-19}$ Gyr$^{-1}$, potentially hinting at a modest ($\approx$2$\times$) increase in \sSFRCSFH{} between $\sim$2 $L_\mathrm{UV}^\ast$ and $\sim$0.4 $L_\mathrm{UV}^\ast$ at $z\sim7-8$ (see Fig. \ref{fig:sSFR}).

To better test for any UV luminosity dependence on \sSFRCSFH{} at $z\sim7-8$, we also consider the large (N=40) sample of extremely UV-bright (median $\Muv{} = -22.0$) galaxies observed with ALMA as part of the Reionization Era Bright Emission Line Survey (REBELS; \citealt{Bouwens2022_REBELS}).
From similar \textsc{beagle} rest-UV$+$optical SED fits, the $\sim$4 $L_\mathrm{UV}^\ast$ REBELS sample has an inferred median $\sSFRCSFH{} = 9.5^{+2.4}_{-2.0}$ Gyr$^{-1}$ \citep{Topping2022_REBELS}, which is nearly an order of magnitude lower than that of our CEERS sample (Fig. \ref{fig:sSFR}).
Given the availability of ALMA data, the CSFH sSFRs of the REBELS objects were also determined using unobscured$+$obscured SFRs computed from the UV and IR luminosities (and the same CSFH stellar masses).
While this approach results in a slightly higher median \sSFRCSFH{} for the REBELS sample ($18^{+7}_{-5}$ Gyr$^{-1}$; \citealt{Topping2022_REBELS}), the data remain consistent with an approximately 5$\times$ higher typical sSFR at $\sim$0.4 $L_\mathrm{UV}^\ast$ relative to $\sim$4 $L_\mathrm{UV}^\ast$ (Fig. \ref{fig:sSFR}).
It may be possible that the rest-UV$+$optical SED fits also significantly underestimate the \sSFRCSFH{} for a subset of the CEERS galaxies, particularly those with very red UV slopes ($\beta \sim -1$; \S\ref{sec:dusty}).
Such a scenario would only boost the inferred median \sSFRCSFH{} of our CEERS sample, thereby strengthening our conclusions of a strong correlation with \Muv{}.
However, the typically bluer UV-slopes of fainter $z\sim7-8$ galaxies suggest that obscured star formation likely plays a less important role in for the CEERS sample as a whole.

The strong UV luminosity correlation with \sSFRCSFH{} described above implies that the light emerging from UV-faint $z\sim7-8$ galaxies is generally more heavily dominated by recently-formed OB stars relative to UV-bright systems at the same epoch.
This in turn may suggest that fainter, more numerous reionization-era galaxies are often more efficient producers of LyC photons.
To explore this, we consider the inferred ionizing photon production efficiencies (\xiion{}) among the CEERS and COSMOS samples, where \xiion{} is defined as the rate at which LyC photons are produced ($\dot{N}_\mathrm{ion}$) divided by the intrinsic stellar UV continuum luminosity at 1500 \AA{} (i.e. that corrected for both dust attenuation and nebular continuum emission) and assuming an escape fraction of $f_\mathrm{esc} = 0$.
From our fiducial \textsc{beagle} fits, we find a very high median \xiion{} among the CEERS galaxies (\logxiion{} = 25.71$^{+0.02}_{-0.01}$) which is $\approx$0.1 dex higher than that inferred from the brighter COSMOS systems (\logxiion{} = 25.61$^{+0.08}_{-0.02}$; see Fig. \ref{fig:sSFR}).
Moreover, we infer that 21\% of the CEERS galaxies possess extremely high \xiion{} ($>10^{25.75}$ erg$^{-1}$ Hz) compared to only 6\% in the COSMOS sample.
The implied increase in \xiion{} towards lower luminosities may suggest that UV-faint galaxies contribute more to reionization than predicted in models that assume \xiion{} is independent of \Muv{} \citep{Naidu2020}.
Larger \xiion{} values at lower luminosities would also alleviate requirements for a high \fesc{} among the numerous, faint galaxy population in order to reach the LyC photon budget necessary for reionization.
Nonetheless, the very blue UV slopes ($\beta \lesssim -2$) of typical sub-$L_\mathrm{UV}^\ast$ reionization-era galaxies suggest these systems may often be efficient leakers of LyC photons \citep[e.g.][]{Chisholm2022}.

\begin{figure}
\includegraphics{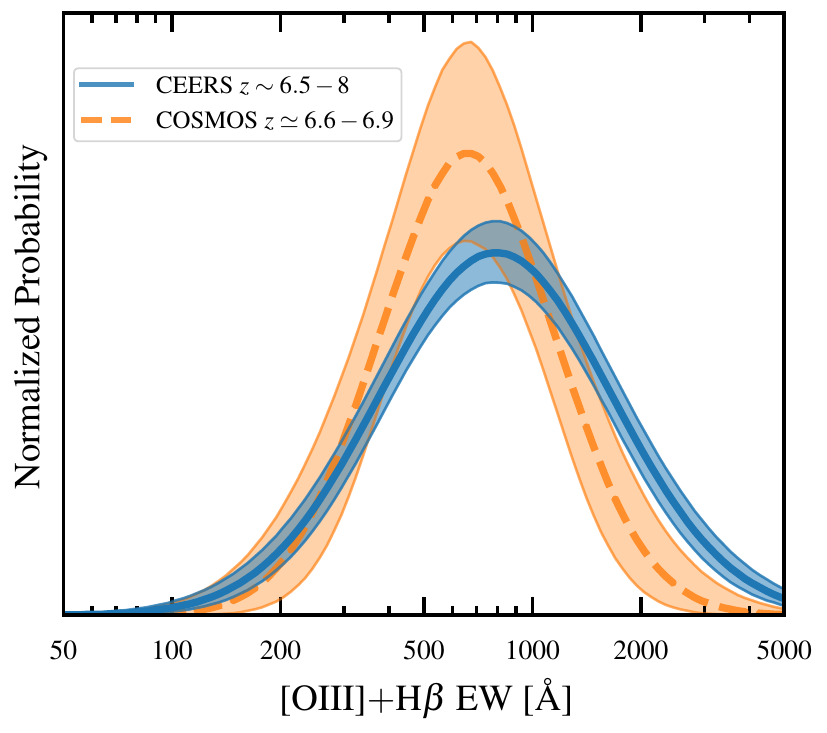}
\caption{The inferred log-normal \OIIIHb{} EW distributions among our relatively faint ($\sim$0.4 $L_\mathrm{UV}^\ast$) CEERS sample (blue) as well as the UV-bright ($\sim$2 $L_\mathrm{UV}^\ast$) COSMOS sample (orange; \citealt{Endsley2021_OIII,Whitler2023_z7}). We find slightly larger EWs among the CEERS sample (median = $780^{+70}_{-50}$ \AA{}) relative to COSMOS (650$^{+110}_{-90}$ \AA{}).}
\label{fig:EWdistns}
\end{figure}

Having investigated the dependence of CSFH sSFR on UV luminosity at $z\sim7-8$, we now turn our attention to the \OIIIHb{} EWs which we expect to be closely tied to \sSFRCSFH{} \citep[e.g.][]{Tang2019}.
For this analysis, we infer the distribution of \OIIIHb{} EWs among our $z\sim6.5-8$ CEERS sample and compare to that of the UV-bright $z\simeq6.6-6.9$ COSMOS sample described above \citep{Endsley2021_OIII,Whitler2023_z7}.
Following the approach of \citet{Endsley2021_OIII}, the EW distribution of each sample is assumed to follow a log-normal shape and is constrained by the \textsc{beagle} posterior probability distribution of all input galaxies adopting the Bayesian formalism of \citet{Boyett2022_OIII}.
We find that the log-normal \OIIIHb{} EW distribution of our CEERS sample is described by a median EW of $780^{+70}_{-50}$ \AA{} and a standard deviation of $0.32^{+0.02}_{-0.03}$ dex (see Fig. \ref{fig:EWdistns}). 
This EW distribution is consistent with the first \JWST{} spectroscopic EW measurements of individual sub-$L_\mathrm{UV}^\ast$ $z\sim7-8$ galaxies \citep[e.g.][]{Schaerer2022}.

Given the strong decrease in \sSFRCSFH{} towards brighter \Muv{} found above (Fig. \ref{fig:sSFR}), we might expect to see evidence of considerably lower \OIIIHb{} EWs among the COSMOS galaxies.
However, the data are consistent with only a slight difference between the COSMOS ($\sim$2 $L_\mathrm{UV}^\ast$) and CEERS ($\sim$0.4 $L_\mathrm{UV}^\ast$) EW distributions, with that of the COSMOS galaxies parametrized by a median EW of 650$^{+110}_{-90}$ \AA{} and standard deviation of 0.24$^{+0.06}_{-0.05}$ dex (Fig. \ref{fig:EWdistns}).
This result is in part driven by the considerable subset (18\%) of our $z\sim6.5-8$ CEERS sample showing SEDs consistent with both high \sSFRCSFH{} ($>$20 Gyr$^{-1}$) and relatively low \OIIIHb{} EWs ($<$600 \AA{}; see \S\ref{sec:weakOIII}), as such a population is not evident from IRAC data in COSMOS.
While our fiducial fits match the photometry of these young, weak \OIIIHb{} emitters by pushing the models to extremely low metallicity (thereby yielding the extended tail in the metallicity distribution shown in Fig. \ref{fig:inferredPropertyHistograms}g), we have shown that alternative solutions (e.g. high $f_\mathrm{esc}$ or a recent strong downturn in SFR) can also provide acceptable fits to the photometry of these objects. 
Given that the \OIIIHb{} EWs are largely constrained by the photometric colors, we do not expect the inferred values to strongly depend on SED model assumptions, though we leave a detailed investigation of possible systematics in the inferred \OIIIHb{} EW distribution to future works.
Upcoming Cycle 1 \JWST{} surveys will clarify the extent to which \sSFRCSFH{} and \OIIIHb{} EW correlate with UV luminosity in the reionization era by extending to wider areas (e.g. COSMOS-Webb, PRIMER), pushing to deeper depths (e.g. JADES, NGDEEP), and delivering spectra (e.g. FRESCO, UNCOVER).
Spectroscopic efforts will simultaneously provide more direct constraints on LyC photon production efficiency, helping build our understanding of the relative contribution of UV-bright vs. UV-faint galaxies to cosmic reionization.

\defcitealias{Labbe2023}{L23}

\begin{figure*}
\includegraphics{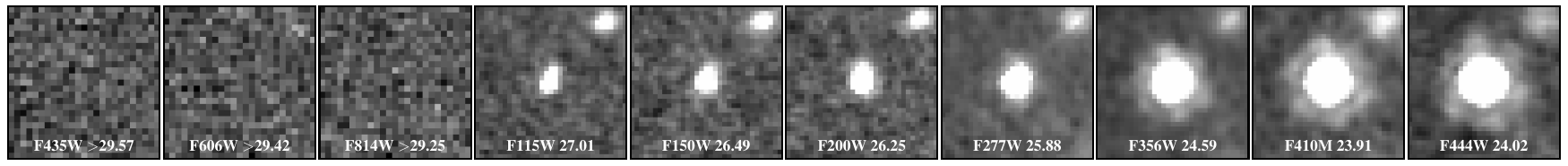}
\includegraphics{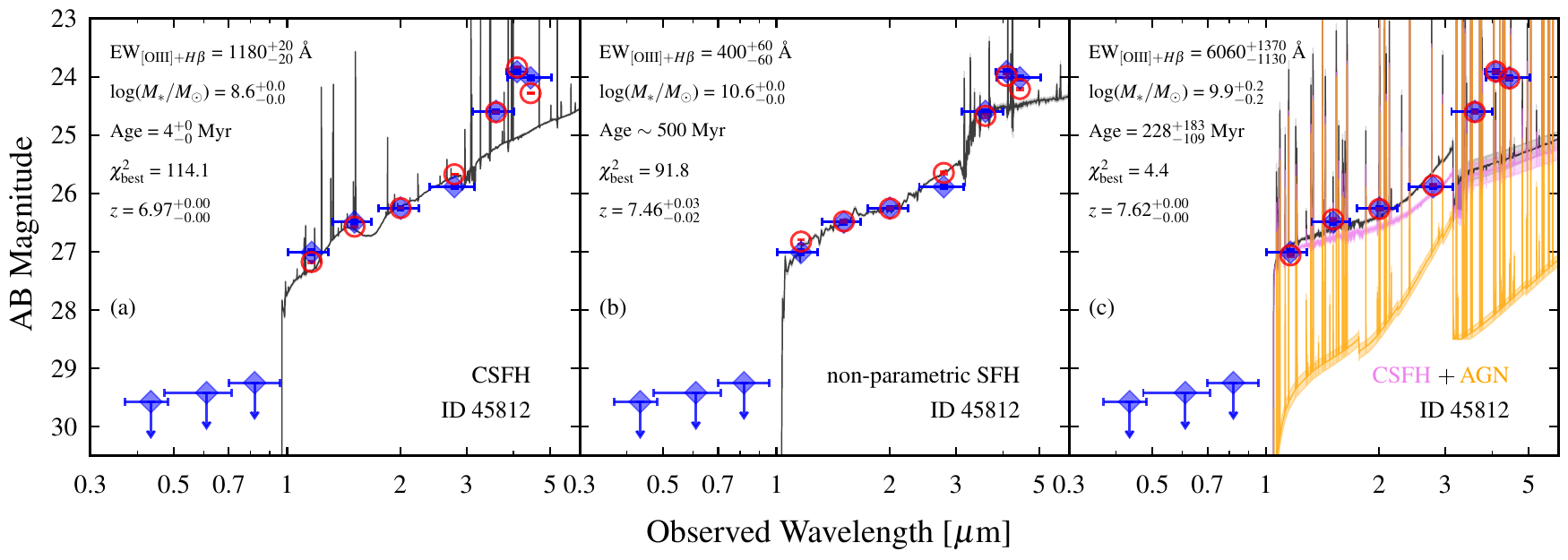}
\caption{Postage stamps (top) and SED fits (bottom) for ID 45812 in our catalog, which was reported to have $M_\ast = 10^{10.9}$ $M_\odot$ in \citet{Labbe2023}. Each stamp is 1$\times$1 arcsec$^2$ and we quote the magnitudes following the format of Fig. \ref{fig:neighborSubtraction}. Bottom panels (a), (b), and (c) show the result from fitting the photometry with a \textsc{beagle} CSFH model, a \textsc{propsector} non-parametric SFH model, and a \textsc{beagle} CSFH$+$AGN model respectively. While by no means conclusive evidence of an AGN in 45812, the LW point-source morphology shown in the postage stamps opens the possibility that the LW excesses are significantly assisted by AGN line emission, consistent with the rightmost model fit. Given the very young (star formation only) CSFH age implied by the LW excesses, we find a very wide ($\approx$2 dex) range of stellar mass solutions for this galaxy depending on model assumptions, though none exceed 5$\times$10$^{10}$ $M_\odot$.} 
\label{fig:AGN}
\end{figure*} 

\subsection{The Stellar Masses of UV-faint \texorpdfstring{$\mathbf{z\sim7-8}$}{z ~ 7 - 8} Galaxies} \label{sec:stellarMasses}

Shortly after NIRCam imaging first became available, several $z\sim7-20$ candidates were identified in small ($\leq$40 arcmin$^2$) fields \citep{Castellano2022_GLASS,Finkelstein2022_z12,Naidu2022_z12,Adams2023,Atek2023,Donnan2023,Harikane2023_z9to17,Labbe2023,Leethochawalit2023,Morishita2023,Whitler2023_z10,Yan2023} some of which were reported to show SEDs suggesting extremely large stellar masses ($\approx10^{10-11}\ M_\odot$; \citealt{Naidu2022_z17,Labbe2023}).
The large abundance of very early massive galaxies implied by these initial studies poses strong challenges for our understanding of baryon accretion onto the largest halos thought to exist at these epochs in context of $\Lambda$CDM \citep{BoylanKolchin2023,Ferrara2023_massivez10,Lovell2023}.
While this potentially points to a failure in our basic understanding of cosmology and/or galaxy formation, it is worth investigating whether considerably lower mass solutions may be viable for some of these reported extremely massive $z\gtrsim7$ candidates.
Here we specifically focus on comparing our stellar mass results with the $z\sim7-11$ sample in \citet{Labbe2023}, hereafter \citetalias{Labbe2023}, since they also utilize the CEERS data and their redshift selection window best overlaps with that considered in this work ($z\sim6.5-8$).

Our fiducial \textsc{beagle} CSFH fits suggest that the NIRCam SEDs of CEERS $z\sim6.5-8$ galaxies are consistent with low stellar masses.
Specifically, we infer a median stellar mass of $M_\ast = 10^{7.9}\ M_\odot$ among our full sample, with only a small fraction (9\%) of objects having $M_\ast > 10^{9}\ M_\odot$ and only one with $M_\ast \approx 10^{10}\ M_\odot$ (ID 1137 in Fig. \ref{fig:ageSEDs}) as illustrated in Figs. \ref{fig:inferredPropertyHistograms}d and \ref{fig:MstarMuv}.
In contrast to our findings, \citetalias{Labbe2023} report the identification of six massive ($M_\ast > 10^{10}\ M_\odot$) $z\sim7-11$ galaxies in the CEERS/NIRCam area, including one object with an extremely high mass ($M_\ast \sim 10^{11}\ M_\odot$).
After cross-matching the \citetalias{Labbe2023} galaxy coordinates with our $z\sim6.5-8$ catalog, we find one $z\sim7$ object which overlaps between the two samples.
This galaxy (ID 45812 in our catalog) is the most massive object in the \citet{Labbe2023} sample with a reported $M_\ast = 10^{10.9}\ M_\odot$.
Upon investigating why the other five \citetalias{Labbe2023} galaxies do not enter our sample, we find that two (including the second most massive object reported in \citetalias{Labbe2023}) fall outside the ACS/F435W footprint and thus were not considered for high-redshift candidate selection in this work.
The final three \citetalias{Labbe2023} galaxies are very faint in F150W and F200W and thus were not identified in our rest-UV detection image by \textsc{source extractor} (\S\ref{sec:photometry}).
\citetalias{Labbe2023} utilized a F277W$+$F356W$+$F444W stack for their detection image which more closely resembles a rest-optical selection, though we emphasize that our rest-UV \textsc{source extractor} catalog does contain the two highest-mass ($M_\ast >10^{10.3}\ M_\odot$) galaxies reported in \citetalias{Labbe2023}.

The NIRCam SED of the \citetalias{Labbe2023} source that overlaps with our $z\sim6.5-8$ selection (ID 45812) is shown in Fig. \ref{fig:AGN}, revealing an extremely red UV slope ($\beta = -0.72^{+0.03}_{-0.02}$) as well as very strong photometric excesses in F356W, F410M, and F444W relative to F277W.
Given the exceptional brightness of this source in the three reddest NIRCam bands ($m=23.9-24.6$), the strength of the photometric excesses is confidently found to vary considerably from band to band.
Because of the high S/N pattern of the LW excesses, our CSFH \textsc{beagle} fits strongly prefer a solution where the SED is dominated by a very young ($\approx$4 Myr) stellar population yielding strong nebular line emission (\OIIIHb{} EW $\approx 1200$ \AA{}; see Fig. \ref{fig:AGN}a).
The nebular line solution allows for more flexibility in the LW colors relative to a solution where the photometric excesses are the result of an approximately power-law rest-optical continuum boosted by an extremely strong Balmer break.
As expected from the very young solution, our CSFH fits yield a relatively low stellar mass of $M_\ast \approx 10^{8.6}\ \Msol{}$, which is about 200$\times$ lower than the $M_\ast \approx 10^{10.9}\ \Msol{}$ solution reported by \citetalias{Labbe2023}.
These \textsc{beagle} CSFH fits demonstrate that relatively low-mass ($\sim4\times10^9\ M_\odot$) solutions are plausible for the highest mass $z\sim7-9$ object reported in \citetalias{Labbe2023} which would alleviate tension with models of galaxy formation \citep{BoylanKolchin2023,Lovell2023}.
Below we comment on models that yield more massive solutions.

It is well known in the literature that CSFH models can significantly underpredict the total stellar mass for galaxies with young SEDs given that these models will only account for stars formed during the recent burst (e.g. \citealt{Carnall2019_SFH,Leja2019,Lower2020}).
Adopting non-parametric SFH models that disfavor rapid changes in the star formation rate will necessarily predict considerably more stellar mass assembly before the onset of the recent burst (see \S\ref{sec:beagle}).
Such non-parametric SFH models have been employed in a number of recent studies focusing on $z\gtrsim7$ galaxies \citep[e.g.][]{Tacchella2022_SFHs,Topping2022_REBELS,Stefanon2023_IRACz10,Tacchella2023_NIRCamNIRSpec,Whitler2023_z10,Whitler2023_z7}, including \citetalias{Labbe2023} which helps explain the extremely high stellar mass they report for ID 45812 ($M_\ast = 10^{10.9}$ $M_\odot$).
As expected, when we fit our photometry for ID 45812 with the non-parametric SFH models described above (following the approach of \citealt{Whitler2023_z7} with a uniform prior formation redshift from $z_\mathrm{form} = 10-30$), we obtain a substantially higher mass solution ($\approx$10$^{10.6}$ $M_\odot$) relative to our \textsc{beagle} CSFH fits (see Fig. \ref{fig:AGN}b). 
However, we note that the non-parametric SFH fit still requires rapid, strong changes in SFR to match the measured photometry for ID 45812.
In the SFH posterior, nearly all of the stellar mass forms between $z_\mathrm{form}\sim20$ and $z\approx11$ (SFR$\approx$200 $M_\odot$ yr$^{-1}$), with very little star formation activity thereafter (SFR$\lesssim$1 $M_\odot$ yr$^{-1}$) until the most recent $\sim$few Myr when star formation ramps back up to power the high S/N excess in F410M.

As can be seen in Fig. \ref{fig:AGN}, neither the CSFH nor non-parametric SFH models are able to precisely reproduce the observed LW colors of ID 45812 (the most massive galaxy reported in \citetalias{Labbe2023}), with each resulting in best-fitting $\chi^2 \approx 100$ across the 10 fitted data points.
Notably, we clearly identify the characteristic 6-pointed diffraction pattern of JWST in the F356W, F410M, and F444W detections of this source (see Fig. \ref{fig:AGN}), indicating that the light dominating the LW bands is coming from a very compact region ($r < 300-400$ pc).
While this is by no means conclusive evidence of an AGN in ID 45812, it at least opens the possibility that the LW excesses are significantly assisted by AGN line emission and we consider whether allowing for such emission (in addition to that from star formation) provides sufficient flexibility in model nebular line strengths and line ratios to better match the data.
Utilizing the \textsc{beagle} AGN models \citep{VidalGarcia2022}, we find that including type II narrow-line AGN emission yields a good fit to the measured photometry (best-fitting $\chi^2 = 4$; Fig. \ref{fig:AGN}c).
In this scenario, the substantial nebular line contribution from an AGN alleviates the need for a very recent upturn of star formation activity, thereby allowing for a much older CSFH solution ($\sim$200 Myr) and a corresponding higher mass-to-light ratio.
The resulting inferred stellar mass of $\approx10^{10}$ $M_\odot$ with the SF$+$AGN models falls between that predicted from the (star formation only) CSFH and non-parametric SFH models described above.

We emphasize that the morphological constraints from JWST can only place a weak upper limit on the effective radius of 45812 in the longest-wavelength bands ($r < 300-400$ pc) and it is therefore not at all clear whether the detections in this band are due to emission from a point-source like an AGN or, alternatively, a very compact star-forming region.
Our primary goal here is to simply demonstrate that allowing for more flexibility in line ratios and line strengths from a model including AGN line emission is one possible way to achieve a good fit ($\chi^2 < 10$) to the measured photometry, and that such a fit yields a moderately massive ($\Mstar{} \approx 10^{10}$ $M_\odot$) solution.
Spectroscopic follow-up will be necessary to clarify the origin of the strong long-wavelength excesses seen in 45812.

Should later spectroscopic results suggest that the photometric excesses seen from 45812 are primarily powered by AGN emission lines, we consider what this may imply for the abundance of AGN at the highest redshifts.
From the \textsc{beagle} AGN fit, we infer an accretion luminosity of $\approx10^{46.0}$ erg/s.
Adopting the bolometric AGN luminosity function from \citet{Shen2020}, we estimate that $N \lesssim 0.01$ such luminous AGN are expected to exist at $z\sim6.5-8$ over the 32 arcmin$^2$ area of CEERS covered by our selection. 
Therefore, if 45812 is found to contain such a luminous AGN, it would imply that either this was an exceedingly fortuitous discovery, or these types of AGN are far more common than previously thought in the reionization era.
Again, we emphasize that spectroscopic follow-up of 45812 (and similar photometric candidates identified in recent months over relatively small fields; \citealt{Furtak2022_uncover,Akins2023}) will be crucial to better assess whether these objects have interesting implications for the abundance of moderately luminous AGN in the first Gyr of cosmic history.

To summarize our key conclusions in this sub-section, our results suggest that the photometry of ID 45812 can be reproduced with models that predict relatively moderate stellar mass ($\sim$10$^{9-10}$ $M_\odot$).
Nevertheless, because this object shows signatures of a very young SED (\ageCSFH{}$\approx$4 Myr), we do find a wide ($\approx$2 dex) range of possible stellar mass solutions depending on model assumptions.
This is indicative of the challenges faced when attempting to infer the absolute stellar masses of galaxies with extremely young SEDs, as was discussed in \citet{Whitler2023_z7}.
Fortunately, there are steps that can be taken to help tighten the range of possible stellar masses for $z\gtrsim7$ galaxies with very young SEDs.
Dynamical mass estimates from NIRCam WFSS or NIRSpec IFU observations would clarify whether very massive ($>10^{10}\ M_\odot$) solutions are plausible for any of these systems \citep[e.g.][]{Tang2022,Topping2022_REBELS}, while targeted searches for the most luminous $z\gtrsim10$ galaxies would help constrain any rigorous star formation activity at much earlier epochs \citep[e.g.][]{Castellano2022_GLASS,Harikane2022_HdropsWideArea,Naidu2022_z12,Donnan2023,Harikane2023_z9to17}.

\section{Summary and Future Directions} \label{sec:summary}

In this work, we characterize the star-forming and ionizing properties of 116 UV-faint Lyman-break $z\sim6.5-8$ galaxies in the CEERS ERS data.
This task is made possible by the greatly improved sensitivity, angular resolution, and filter suite of \JWST{}/NIRCam relative to \Spitzer{}/IRAC, resulting in far better constraining power on the demographics of rest-optical SEDs among sub-$L_\mathrm{UV}^\ast$ reionization-era galaxies (see Fig. \ref{fig:HSTSpitzerComparison}).
We infer the physical properties of each CEERS $z\sim6.5-8$ galaxy by fitting their 0.4--5$\mu$m photometry to a suite of star-forming photoionization model SEDs with \textsc{beagle} \citep{Chevallard2016,Gutkin2016}.
Following many previous studies at $z\sim6-8$, we adopt constant star formation history (CSFH) models for our fiducial fits, and we comment on how alternative star formation histories would change our results.
To promote comparisons between independent analyses, we provide an online catalog (see Table \ref{tab:properties}) listing coordinates, flux densities, and inferred physical properties among our sample.
Our main conclusions are summarized below, where we note how future observations can help clarify outstanding questions raised by our results.

\begin{enumerate}
    
    \item The photometric redshifts of galaxies in our sample range between $z=6.27$ and $z=7.97$ with a median of $z=6.82$, while the inferred absolute UV magnitudes fall between $-21.0 \leq \Muv{} \leq -18.8$ with a median of $\Muv{} = -19.5$ (see Fig. \ref{fig:inferredPropertyHistograms}). Adopting a characteristic UV luminosity corresponding to M$_\mathrm{UV}^\ast = -20.5$ \citep[e.g.][]{Bowler2017,Harikane2022_LF}, this implies that the typical galaxy in our sample is 0.4 $L_\mathrm{UV}^\ast$ and 92\% of our sources are classified as sub-$L_\mathrm{UV}^\ast$ systems.
    
    \item The NIRCam data clearly reveal a wide variety of CSFH ages among the sample (Fig. \ref{fig:ageSEDs}). For the bulk of galaxies in our sample, the measured flux density in either F356W, F410M, or F444W is consistent with extrapolating a power-law SED from the rest-UV photometry, implying that their SEDs are dominated by young stellar populations (\ageCSFH{}$\sim$30 Myr). Nonetheless, we do identify a considerable subset of objects ($\approx$30\% of the sample) with significant and nearly equal flux excesses in the three reddest NIRCam bands, implying the presence of a prominent Balmer break consistent with relatively evolved stellar populations (\ageCSFH{}$\sim$100-500 Myr). Another $\approx$30\% of galaxies in our sample show very strong long-wavelength photometric excesses with the extent of the excess varying significantly from band to band, implying contamination from exceptionally high-EW nebular lines (\OIIIHb{} EW$\gtrsim$1500 \AA{}) powered by very young stellar populations (\ageCSFH{}$\lesssim$10 Myr). In context of our fiducial CSFH models, this broad distribution in ages directly implies that sub-$L_\mathrm{UV}^\ast$ reionization-era galaxies have a wide diversity of \sSFRCSFH{} ($\approx$2 Gyr$^{-1}$ to $>$500 Gyr$^{-1}$).

    \item We find that UV-faint ($\sim$0.4 $L_\mathrm{UV}^\ast$) $z\sim6.5-8$ galaxies typically have very large specific star formation rates (median $\sSFRCSFH{} = 103^{+9}_{-28}$ Gyr$^{-1}$; see Fig. \ref{fig:sSFR}), consistent with a scenario in which many of these systems recently experienced a strong upturn in their star formation activity. Combining our results with previous \Spitzer{}/IRAC studies of the UV-bright $z\sim7-8$ galaxy population \citep{Endsley2021_OIII,Topping2022_REBELS,Whitler2023_z7}, we find evidence for a strong ($\approx$5--10$\times$) increase in \sSFRCSFH{} between $\sim$4 $L_\mathrm{UV}^\ast$ and $\sim$0.4 $L_\mathrm{UV}^\ast$ (Fig. \ref{fig:sSFR}). This implies that recently-formed OB stars contribute a larger fraction of the emergent light from fainter $z\sim7-8$ galaxies, resulting in a slight increase in the typical ionizing photon production efficiency in the CEERS sample (median \xiion{} $\approx$ 10$^{25.71}$ erg$^{-1}$ Hz in CEERS) relative to the UV-bright systems (\xiion{} $\approx$ 10$^{25.61}$ erg$^{-1}$ Hz). This $\approx$0.1 dex increase in \xiion{} coupled with the steep faint-end slope of the $z\sim7-8$ UV luminosity function could imply that sub-$L_\mathrm{UV}^\ast$ reionization-era galaxies contribute more to reionization than previously thought. Future \JWST{} surveys will help clarify the relative contribution of bright vs. faint galaxies to cosmic reionization by extending to wider areas (e.g. COSMOS-Webb, PRIMER), pushing to deeper depths (e.g. JADES, NGDEEP), and delivering spectra (e.g. FRESCO, UNCOVER).
    
    \item We infer the \OIIIHb{} EW distribution among UV-faint ($\sim$0.4 $L_\mathrm{UV}^\ast$) $z\sim6.5-8$ galaxies assuming it follows a log-normal function, finding a median EW = 780$^{+70}_{-50}$ \AA{} with a standard deviation of 0.32$^{+0.02}_{-0.03}$ dex (Fig. \ref{fig:EWdistns}). Despite the fact that we find much larger \sSFRCSFH{} among our CEERS galaxies, we find only a slight increase in the \OIIIHb{} EWs relative to more UV-luminous systems ($\sim$2 $L_\mathrm{UV}^\ast$; Fig. \ref{fig:EWdistns}). This is largely due to the substantial fraction ($\approx$18\%) of our CEERS galaxies that show SEDs consistent with high \sSFRCSFH{} ($>$20 Gyr$^{-1}$) yet relatively weak \OIIIHb{} (EW$<$600 \AA{}; see Fig. \ref{fig:youngWeakOIIISEDs}). Such a population is not evident among brighter $z\sim7-8$ galaxies from IRAC data, nor among existing samples at $z\sim2$ covering similar \OIIIHb{} EWs \citep{Tang2019}. In context of our fiducial CSFH \textsc{beagle} models, young SEDs with weak \OIIIHb{} are often reproduced with extremely low metallicities ($\lesssim$3\% $Z_\odot$; see Fig. \ref{fig:age_EW_metallicity}) which greatly suppresses the [OIII] emission. However, we also demonstrate that high ionizing photon escape fractions ($f_\mathrm{esc,HII} \gtrsim 0.5$) or a recent ($\sim$10 Myr) sharp decline in star formation rate can also reproduce the SEDs for a subset of our galaxies (Fig. \ref{fig:youngWeakOIIISEDs_otherExplanations}). Future spectroscopic observations will be necessary to not only confirm the existence of this seemingly important class of faint reionization-era galaxies, but also to better determine the physical origin of their weak \OIIIHb{}.
    
    \item As expected from previous studies utilizing deep \HST{}/WFC3 data, we find that our UV-faint CEERS galaxies generally exhibit very blue UV slopes (median $\beta = -2.1$; Fig. \ref{fig:inferredPropertyHistograms}). Nonetheless, we identify three galaxies with high S/N ($\gtrsim$20$\sigma$) short-wavelength NIRCam photometry indicating very red UV slopes ($-1.4\leq\beta\leq-0.7$), as well as long-wavelength photometric excesses implying strong \OIIIHb{} (EW$>$800 \AA{}) at $z\sim7$ (Fig. \ref{fig:redEELGseds}). While all three systems are classified as $\approx$0.3--0.5 $L_\mathrm{UV}^\ast$ from the data, their very red UV slopes imply that they would appear as far more luminous galaxies in the absence of dust ($\approx$1--4 $L_\mathrm{UV}^\ast$). The fact that we identify three such objects in CEERS suggests that there remains a significant tail of heavily dust-reddened $z\sim7$ galaxies down to faint M$_\mathrm{UV}$. The \textsc{beagle} fits imply that $\approx$70--90\% of their star formation is obscured which in total contributes 19\% of all (unobscured$+$obscured) stellar mass growth inferred among our CEERS sample, even though these objects make up only 3\% of the sample by number. Deep far-infrared follow-up will be critical to assess the extent of obscured star formation in UV-faint yet very red $z\sim7$ galaxies identified from deep \JWST{} data.
    
    \item Our fiducial CSFH \textsc{beagle} fits imply that the NIRCam SEDs of UV-faint ($\Muv{} \sim -19.5$) $z\sim6.5-8$ galaxies are consistent with low stellar masses (median $M_\ast = 10^{7.9}\ M_\odot$), and only one object in our sample has \MstarCSFH{}$\approx$10$^{10}$ $M_\odot$. We investigate one $z\sim7$ object in our sample that was reported to have an extremely large stellar mass ($10^{10.9}$ $M_\odot$) in a previous work \citep{Labbe2023} which challenges models of galaxy formation \citep{BoylanKolchin2023,Ferrara2023_massivez10,Lovell2023}. This galaxy shows an extremely red UV slope ($\beta = -0.7$) with a high S/N LW photometric excess pattern that implies strong nebular line emission and hence a very young (\ageCSFH{}$\approx$4 Myr) SED. Because the recent burst may be outshining an older stellar population, we find a very wide range of plausible stellar mass solutions for this galaxy ($\approx$10$^{8.6}$ $M_\odot$ to $\approx$10$^{10.6}$ $M_\odot$) depending on the assumed star formation history (Fig. \ref{fig:AGN}a,b). Given the extreme LW brightness of this object, we are able to clearly identify the characteristic 6-pointed diffraction pattern of JWST in the F356W, F410M, and F444W detections (see Fig. \ref{fig:AGN}). While this is by no means conclusive evidence of an AGN (spectroscopic follow-up is necessary), it at least opens the possibility that the LW excesses are significantly assisted by AGN line emission and we find that the inclusion of such models yield a good fit to the measured photometry with $M_\ast \approx 10^{10}$ $M_\odot$ (Fig. \ref{fig:AGN}c). Dynamical mass estimates from future spectroscopic observations will help clarify the range of plausible stellar mass solutions for $z\gtrsim7$ galaxies with very young SEDs \citep[e.g.][]{Tang2022,Topping2022_REBELS}. 
   
\end{enumerate}

\section*{Acknowledgements}

The authors thank the CEERS team for their hard work designing and executing the CEERS program. In addition, we thank the entire JWST team for their work designing, building, commissioning, and developing tools for the telescope. Finally, we thank the anonymous referree for helpful constructive comments, Mengtao Tang for providing information on their $z\sim2$ EELG sample, Gabe Brammer for providing the ACS imaging mosaics of the EGS field as part of CHArGE program, as well as Jacopo Chevallard for access to the \textsc{beagle} tool used for much of our SED fitting analysis.

RE and DPS acknowledge funding from JWST/NIRCam contract to the University of Arizona, NAS5-02015.
DPS acknowledges support from the National Science Foundation through the grant AST-2109066.
LW acknowledges support from the National Science Foundation Graduate Research Fellowship under Grant No. DGE-2137419.

This material is based in part upon High Performance Computing (HPC) resources supported by the University of Arizona TRIF, UITS, and Research, Innovation, and Impact (RII) and maintained by the UArizona Research Technologies department.

This work made use of the following software: \textsc{numpy} \citep{harris2020_numpy}; \textsc{matplotlib} \citep{Hunter2007_matplotlib}; \textsc{scipy} \citep{Virtanen2020_SciPy}; \textsc{astropy}\footnote{\url{https://www.astropy.org/}}, a community-developed core Python package for Astronomy \citep{astropy:2013, astropy:2018}; \textsc{Source Extractor} \citep{Bertin1996} via \textsc{sep} \citep{Barbary2016_sep}; \textsc{photutils} \citep{Bradley2022_photutils}; \textsc{beagle} \citep{Chevallard2016}; \textsc{multinest} \citep{Feroz2008,Feroz2009};  \textsc{prospector} \cite{Johnson2021}; \textsc{dynesty} \citep{Speagle2020}; \textsc{sedpy} \citep{johnson_sedpy}; and \textsc{fsps} \citep{Conroy2009_FSPS,Conroy2010_FSPS} via \textsc{python-fsps} \citep{johnson_python_fsps}.

\section*{Data Availability}

The \HST{}/ACS and \JWST{}/NIRCam images used in this work are available through the Mikulski Archive for Space Telescopes (\url{https://mast.stsci.edu/}). 
We provide an online catalog listing the coordinates, magnitudes, and inferred physical properties of our $z\sim6.5-8$ galaxy sample in CEERS.
Additional data products will be made available upon reasonable request to the corresponding author.




\bibliographystyle{mnras}
\bibliography{main} 



\appendix
 

\bsp	
\label{lastpage}
\end{document}